\documentclass[
 reprint, 
 pra,
 amsmath,amssymb,
 aps, 
 superscriptaddress, nofootinbib
]{revtex4-2}

\usepackage{graphicx}
\usepackage{dcolumn}
\usepackage{bm}
\usepackage[colorlinks, allcolors=blue]{hyperref}
\usepackage[all]{hypcap}
\usepackage{amsmath}
\usepackage{caption}
\usepackage{subcaption}
\usepackage[normalem]{ulem}
\usepackage{stmaryrd}
\usepackage{bbold}
\usepackage{comment}
\usepackage{xcolor}

\newcommand{\ket}[1]{\vert #1\rangle}
\newcommand{\bra}[1]{\langle #1\vert}

\newcommand{\pref}[2]{\hyperref[#1]{\ref{#1}(#2)}}

\newtheorem{definition}{Definition}

\newcommand{\hide}[1]{}

\begin{document}
\title{Decoding Quantum Low Density Parity Check Codes with Diffusion}

\def\urbana{
The Anthony J. Leggett Institute for Condensed Matter Theory and IQUIST and NCSA Center for Artificial Intelligence Innovation and Department of Physics, University of Illinois at Urbana-Champaign, Urbana, Illinois, USA}
\def\zurich{
Institute for Theoretical Physics, ETH Z\"{u}rich, Switzerland}
\author{Zejun Liu}
\affiliation{\urbana}
\author{Anqi Gong}
\affiliation{\zurich}
\author{Bryan K. Clark}
\affiliation{\urbana}

\begin{abstract}
An efficient decoder is essential for quantum error correction, and data-driven neural decoders have emerged as promising, flexible solutions. 
Here, we introduce a diffusion (DF) model framework to infer logical errors from syndrome measurements in quantum low-density parity-check (qLDPC) codes. 
Using the bivariate bicycle code with realistic circuit-level noise, we show that masked diffusion decoders are more accurate, often faster on average, and always faster in the worst case than other state-of-the-art decoders, including belief propagation with ordered statistics decoding (BP-OSD) and autoregressive neural decoders.  We show that by using fewer diffusion steps during inference one can gain significant speed at minimal cost in accuracy.  
By examining the factored attention from our trained neural network we find that, despite being trained solely on paired samples of syndrome–logical errors, this diffusion decoder learns the structure of the quantum codes.  
We also compare both masked and continuous diffusion decoders on code-capacity noise models, finding that masked diffusion decoders scale better than continuous diffusion decoders. 
\end{abstract}
\maketitle
\tableofcontents
\section{Introduction}
In quantum computing, the decoherence of qubits over time necessitates the use of quantum error correction (QEC) to preserve their logical state~\cite{deutsch1985quantum,shor1995scheme}. QEC works by encoding a number of logical qubits into the Hilbert space of a larger number of physical qubits and then measures observables (check operators) on some of these qubits, generating syndrome information from which one aims to infer the underlying errors.
A \textit{decoder} then must map the error syndrome to the occurred physical or logical error, which is then corrected. Decoders need to be both accurate in determining the correct error that occurred as well as fast to avoid bottlenecking the quantum computation and avoiding further decoherence before they can be corrected~\cite{Terhal2013Quantum}.

While many different error correcting codes exist, quantum low-density parity-check (qLDPC) codes have recently emerged as a new standard for QEC with higher encoding rates and better scaling of code distance~\cite{mackay2004sparse,tillich2013quantum,panteleev2021degenerate,panteleev2021quantum,breuckmann2021balanced} than previous approaches such as the surface codes~\cite{dennis2002topological,google2023suppressing,google2025quantum,bluvstein2024logical}. 
Unlike surface code constructions, qLDPC codes encode many $(k>1)$ logical qubits into $n$ physical qubits simultaneously, and thus the decoding process also requires simultaneous predictions over these $(k>1)$ logical qubits.   
Recent research efforts include various constructions of this family of quantum codes~\cite{breuckmann2021quantum,panteleev2021degenerate, bravyi2024high}, experimental demonstration of small-scale qLDPC codes~\cite{wang2025demonstration}, and the design of decoding algorithms~\cite{panteleev2021degenerate,muller2025improved,gong2024toward,yao2024belief,iolius2024almost,hillmann2024localized,ott2025decision,blue2025machine,cao2025generative,wu2025MWPF,hu2025efficient}.

Belief propagation with ordered statistics decoding (BP-OSD), where the BP stage is iterative and OSD involves finding solution of linear equations using Gaussian elimination (GE), has been considered as a standard for decoding qLDPC codes~\cite{panteleev2021degenerate, roffe2020decoding}. Despite being a universal solution, it is not clear that the worst-case decoding time can meet the real-time decoding requirements for fault tolerance. Some efforts have attempted to overcome this bottleneck through a parallelized version of OSD by dividing the decoding problem into small clusters/instances \cite{wolanski2024ambiguity,hillmann2024localized}. Other works, in contrast, try to get rid of the Gaussian elimination part by running BP repeatedly on a modified Tanner graph \cite{yao2024belief,gong2024toward,iolius2024almost,ott2025decision,muller2025improved}. 

In this work, we improve upon an alternative approach which uses neural decoders \cite{blue2025machine,cao2025generative,bausch2024learning,liu2019neural,gong2024graph,maan2025machine} in lieu of BP. Neural decoders offer a data-driven and versatile alternative, since in a real and complicated system, they have the potential to capture implicit correlations that would otherwise be too difficult to specify. 
Training and inference can be accelerated on GPUs for fast prototyping of the algorithms. 
A prominent feature of neural decoders is that, without a post-processing stage, the decoding time is almost the same for any given syndrome (there is no notion of \emph{worst-case} time); this is in stark contrast to e.g., BP-OSD. For a well-trained and reasonably-sized model, the blooming field of custom-made neural accelerators could open up the possibility of further improving inference time.

\begin{figure}
\centering
\includegraphics[width=0.5\textwidth]{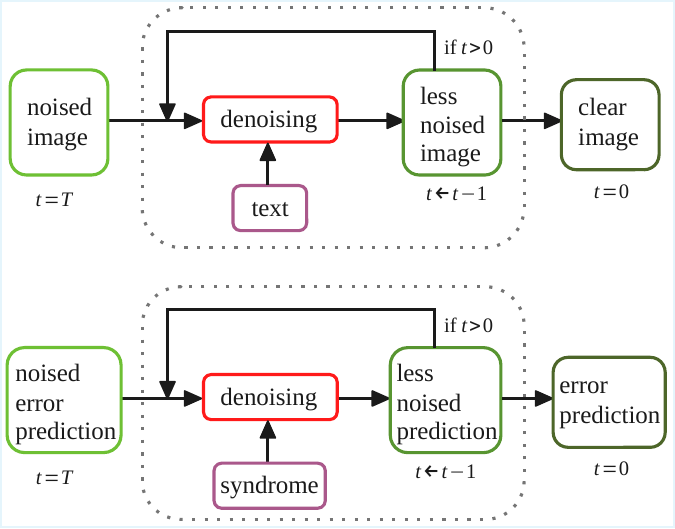}
\caption{Illustration of diffusion decoder that predicts errors for QEC conditioned on syndromes (bottom), in comparison with the image generation using diffusion model conditioned on text description for the image (top).}
\label{fig:decoder}
\end{figure}

Previous neural decoders employ auto-regression (AR) models to predict the logical observables sequentially~\cite{cao2025generative,blue2025machine}. 
We introduce the diffusion (DF) decoder as an alternative neural decoding framework, which allows for parallel prediction of logical observables due to the diffusion model paradigm~\cite{ho2020denoising,chang2023muse,sahoo2024simple,shi2024simplified,kim2025train,nie2025large} we follow (Fig.~\ref{fig:decoder}).
This approach provides a more symmetric treatment of the logical errors as there is no inherent ordering and is therefore well-aligned with the symmetry from the qLDPC code design. 
The denoising process in the DF decoder reveals logical errors based on confidence scores from the trained neural network, sharing similarity to
algorithms that make use of the posterior probability from iterative belief propagation~\cite{yao2024belief,gong2024toward}. Thus, this framework allows us to take advantage of the strengths of data-driven learning and remain aligned with established decoding intuitions.
Our results also give discernible improvements in accuracy and decoding time using fewer parameters.

\subsection{qLDPC codes and error models}
A quantum circuit consists of a series of qubits that are acted on by gates and measurements. 
During the operation of quantum circuits, the interaction with the environment induces erroneous operations on the physical qubits.

In QEC with stabilizer codes, the logical state is encoded as a simultaneous eigenstate of a set of pairwise commuting Pauli strings called check operators.
The physical qubits decohere continuously as a function of time; 
measurements of the check operators then project the continuous decoherence into a set of discrete errors which act as the probabilistic application of a Pauli string onto the state.  An error channel is defined by a distribution over these probabilities.   
The outcomes from the measurements of the check operators give the syndrome information about the physical/logical error on the post-measured state.
A prototypical example of an error channel is the depolarizing channel acting on each physical qubit independently:
\begin{equation}
\begin{split}
\label{eq:depolar}
\mathcal{E}_q(\ket{\psi}\bra{\psi})&=(1-p)\ket{\psi}\bra{\psi}\\&+\frac{p}{3}X_q\ket{\psi}\bra{\psi}X_q+\frac{p}{3}Y_q\ket{\psi}\bra{\psi}Y_q+\frac{p}{3}Z_q\ket{\psi}\bra{\psi}Z_q    
\end{split}
\end{equation}
where $\{X_q, Y_q, Z_q\}$ are the Pauli operators acting on qubit $q$. Quantum LDPC codes, as an example of stabilizer codes, are attractive because the number of qubits involved in each check operator measurement and the number of check operators each qubit participates in are bounded by a constant or grow slowly with respect to the physical qubit number~\cite{mackay2004sparse,tillich2013quantum,panteleev2021degenerate,panteleev2021quantum,breuckmann2021balanced,breuckmann2021quantum}.
A conventional approach is to measure the syndromes with the bare-ancilla approach where one initializes additional qubits in $\ket{0}$ or $\ket{+}$, copies the error from the data qubits onto them, and then measures them in the $Z$ or $X$ basis to obtain the syndrome.  Because the checks in the qLDPC codes are low-weight, this keeps the depth of the syndrome extraction circuit low and alleviates the introduction of additional errors. 

There are two standard noise models to gauge the noise resilience of quantum codes: the code-capacity and circuit-level noise models~\cite{dennis2002topological,Terhal2013Quantum}. 
In the code capacity setting, there is a simplified assumption that errors do not happen while measuring the check operators; it offers a simple platform for estimating the performance and threshold under certain decoders for a family of quantum codes.
Circuit-level noise, which is the primary focus of this work, is a more realistic model where all operations (one and two-qubit gates, measurements) and idle qubits are subject to probabilistic errors. 
In this latter case, since the measurement of check operators might be unreliable, usually one repeats the measurements for multiple rounds, so that enough information can be provided to the decoder for more reliable inference of errors.
In the course of repeated measurements, a single error can propagate into future rounds and flip many measurement results. To ease the decoding problem (sparsify the decoding graph), the syndrome is defined as the XOR between measurement outcomes of the same check from contiguous rounds \cite{bravyi2024high}. Due to the repeated measurements, the circuit-level noise model has a much larger syndrome size; the number of possible error events is also orders of magnitude larger compared to the code-capacity setting.

Formally, in both code-capacity and circuit-level noise setups, the syndrome $\bm{s}\in\{0,1\}^{n_s}$ is related to the physical error $\bm{e}\in\{0,1\}^{n_e}$ (where $1$ indicates this error occurs in the quantum system and $0$ indicates otherwise) via a binary parity check matrix (PCM) $H\in\{0,1\}^{n_s\times n_e}$ by $H\bm{e}=\bm{s}$, where $H_{ij}=1$ indicates that physical error event $j$ triggers the syndrome $i$ in the absence of other error events. Note that there are many possible errors $\bm{e}$ which could give the same $\bm{s}$. Similarly, a logical matrix $L\in\{0,1\}^{n_l\times n_e}$ can be defined for the logical operators, where $L_{ij}=1$ indicates error event $j$ triggers the logical error $i$ in the absence of other physical error events.  
The goal of decoding is to quickly determine the most probable $\bm{e}$ given $\bm{s}$ or alternatively the most probable logical errors $\bm{l}=L\bm{e}$~\cite{Terhal2013Quantum}. 
We can formulate the decoding problem as follows:
\begin{definition}[Decoding problem]
Given a syndrome $\bm{s}$ sampled from 
\begin{equation}
p(\bm{s}) = \sum_{\bm{e}\in\{\bm{e}|H\bm{e}=\bm{s}\}}\Pr[\bm{e}]    
\end{equation}
where
\begin{equation}
\Pr[\bm{e}]=\prod_{j=1}^{n_e}\Pr[e_j] 
\end{equation}
with $\Pr[e_j=1]=p_j$ as the probability for error event $j$ to happen, find the most probable $\hat{\bm{e}}$ from the conditional probability
\begin{equation}
p(\bm{e|\bm{s}})=\frac{p(\bm{s}|\bm{e})p(\bm{e})}{p(\bm{s})}=\frac{\delta_{H\bm{e}=\bm{s}}\Pr[\bm{e}]}{\sum_{\bm{e}\in\{\bm{e}|H\bm{e}=\bm{s}\}}\Pr[\bm{e}]}    
\end{equation}
i.e.,
\begin{equation}
\hat{\bm{e}}=\arg\max_{\bm{e}} p(\bm{e}|\bm{s})       
\end{equation}
or the most probable logical error $\hat{\bm{l}}$ from the conditional probability
\begin{equation}
p(\bm{l}|\bm{s})=\frac{p(\bm{l},\bm{s})}{p(\bm{s})}=\frac{\sum_{\bm{e}\in\{\bm{e}|H\bm{e}=\bm{s},L\bm{e}=\bm{l}\}}\Pr[\bm{e}]}{\sum_{\bm{e}\in\{\bm{e}|H\bm{e}=\bm{s}\}}\Pr[\bm{e}]}   
\end{equation}
i.e.,
\begin{equation}
\hat{\bm{l}} = \arg\max_{\bm{l}} p(\bm{l}|\bm{s}).       
\end{equation}
\end{definition}

\subsection{Diffusion decoders}
Evaluating $p(\bm{e}|\bm{s})$ or $p(\bm{l}|\bm{s})$ is generally intractable due to the exponential size of the error space. Hence, neural decoders are introduced to provide an accessible surrogate $q_{\bm{\theta}}(\bm{e}|\bm{s})$ or $q_{\bm{\theta}}(\bm{l}|\bm{s})$ 
which is learned from the
data $\left(\bm{s}, \bm{e}\right)$ or $\left(\bm{s}, \bm{l}\right)$ and from which the most likely error can be generated as 
\begin{equation}
\hat{\bm{e}}(\bm{s})=\arg\max_{\bm{e}} q_{\bm{\theta}}(\bm{e}|\bm{s}) \text{ or }  
\hat{\bm{l}}(\bm{s})=\arg\max_{\bm{l}} q_{\bm{\theta}}(\bm{l}|\bm{s})     
\end{equation}
Since typically $n_e\gg n_l$, especially for circuit-level noise decoding, it is easier for neural decoders to model $p(\bm{l}|\bm{s})$ than to model $p(\bm{e}|\bm{s})$. Predicting the logical error is sufficient for quantum memory experiments~\cite{dennis2002topological,bausch2024learning,google2023suppressing,google2025quantum,bluvstein2024logical} or Pauli-based computation~\cite{bravyi2016trading,blue2025machine,zhou2024algorithmicFT}. To evaluate the performance of this kind of decoder, the logical error rate (LER) $p_L$ is defined as 
\begin{equation}
p_L=\frac{\text{number of samples } \hat{\bm{l}}\neq\bm{l}}{\text{number of samples } (\bm{e},\bm{s},\bm{l})}.
\end{equation}

The DF decoders generate the logical error $\bm{l}$ through a Markov chain, where the variational transition probability $q_{\bm{\theta}}(\bm{l}_{t-1}|\bm{l}_t,\bm{s})$ ($t=1,2,\dots,T$) is computed from a neural network parameterized by $\bm{\theta}$ and conditioned on both the syndrome $\bm{s}$ and the logical error prediction $\bm{l}_t$ at step $t$.
Notice that, in contrast to standard diffusion generative models in which $\bm{l}_{t-1}$ is randomly sampled from the transition probability $q_{\bm{\theta}}(\bm{l}_{t-1}|\bm{l}_t,\bm{s})$, here our goal is the most probable final state $\bm{l}\equiv\bm{l}_0$
which we obtain greedily at each step by  
\begin{equation}
\label{eq:reverse_markov}
\hat{\bm{l}}_{t-1}=\arg\max_{\bm{l}_{t-1}} q_{\bm{\theta}}(\bm{l}_{t-1}|\bm{l}_{t},\bm{s}) 
\end{equation}
and iterating this process to reach an approximate maximal $\bm{l}_0$. 
This approach trades off accuracy for speed and it remains an open question how closely this heuristic maximization matches the actual maximum and whether there exist better protocols.

We consider both the continuous diffusion model and masked diffusion model in this work. In the continuous DF model, the forward diffusion process is to add independent random Gaussian noises to each element in $\bm{l}$ and arrive at a normal distribution for $\bm{l}_T$~\cite{chen2022analog,choukroun2022denoising}. 
The corresponding reverse process has Gaussian transition probability $q_{\bm{\theta}}(\bm{l}_{t-1}|\bm{l}_t,\bm{s})$ with a neural-network generated mean $\bm{\mu}_{\bm{\theta}}(\bm{l}_t,\bm{s},t)$ which is the maximal to which $\bm{l}_{t-1}$ is updated. In the masked DF model, the forward process is to independently mask each element in $\bm{l}$ with a predetermined probability at each time step with $\bm{l}_T$ being fully masked. Consequently, the reverse unmasking process gives the probabilities for each masked element in $\bm{l}_t$ to take value 0/1, and our heuristic for choosing the maximal is then to unmask the elements whose probabilities are most confident (i.e., those closest to 0 or 1) at each step~\cite{austin2021structured,nie2025large}.

The masked DF model shares some similarities with the AR models~\cite{austin2021structured,nie2025large}. Operationally, for AR decoders, a partially masked $\bm{l}$ is fed into the neural network to predict the next logical error bit sequentially~\cite{blue2025machine,cao2025generative}; for the masked DF decoder, the neural network also receives a partially masked $\bm{l}$ and some of the masked bits are assigned as 0/1 based on the output probabilities. On the other hand, there are non-trivial differences between these two frameworks. Instead of predicting one bit at a time, DF decoder outputs the probabilities for all the masked bits, and has the flexibility to choose most confident bits to predict. 
The training protocols of the AR and masked DF decoders are different: while the training samples $\bm{l}$ for AR decoders are masked causally (i.e., all bits after a bit position are masked), the training samples for masked DF decoders are masked randomly without causality. Hence, BERT~\cite{devlin2019bert,vaswani2017attention} becomes a natural choice as the core neural network architecture for masked DF decoders (see Sec.~\ref{sec:NNdetail}).

Compared to previous AR decoders~\cite{blue2025machine}, our neural networks are designed with simpler structures with much fewer parameters while maintaining the quality of performance. In particular, inside the BERT architecture, we use the factored attention~\cite{viteritti2023transformer} instead of the standard attention~\cite{vaswani2017attention}: while the standard attention computes the attention matrix from the query and key matrices that come from the linear mappings of the input, the factored attention directly assigns the attention matrix as learnable parameters that are independent of the input. This change reduces the times of linear mappings and hence the running time and parameter number of the neural network. 

The codes to reproduce our main results are provided at Ref.~\cite{zenodo_code}.

\section{Results}
\subsection{Decoding circuit-level error}

\begin{figure*}
\centering
\includegraphics[width=\textwidth]{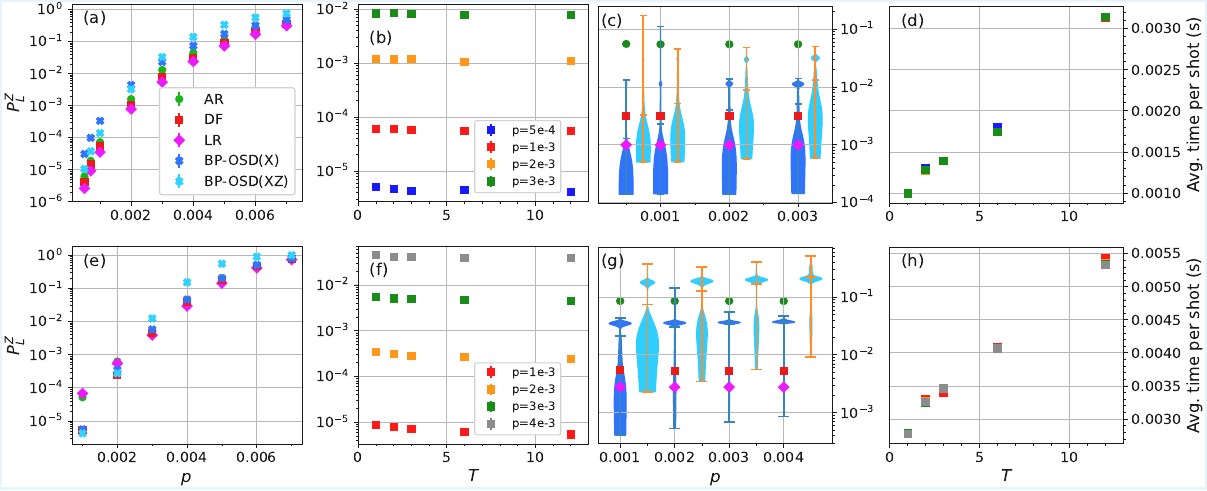}
\caption{Circuit-level noise decoding results on (a)-(d) $\llbracket 72,12,6\rrbracket$ and (e)-(h) $\llbracket144,12,12\rrbracket$ BB codes. (a)/(e) Comparison of LER among different decoders: BP-OSD(X) (using only X syndromes) and BP-OSD(XZ) (using both X and Z syndromes for decoding) using CUDA-Q~\cite{NVIDIA_Corporation_CUDA-QX}, AR decoder from Ref.~\cite{blue2025machine}, masked DF decoders trained with $T=n_l=k$ (DF) and the masked DF decoder trained with $T=1$ (LR). (c)/(g) Comparison of decoding time per sample (latency) among these decoders: the BP-OSD(X(deep blue)/XZ(light blue)) data show the distribution from 10,000 samples at each $p$, with BP-OSD(XZ) data shifted right for better visualization.
Dependence of (b)/(f) LER and (d)/(h) average decoding time from masked DF decoder on $T$ during inference: all the data come from the same DF decoder trained with $T=n_l$. The latency data are obtained from an NVIDIA A100 GPU with batch size of one, the same setting as the AR decoder from Ref.~\cite{blue2025machine}.}
\label{fig:circlevel}
\end{figure*}

We apply the masked DF decoder to memory experiments under a circuit-level noise model for the $\llbracket n,k,d\rrbracket=\llbracket 72,12,6\rrbracket$ and $\llbracket 144,12,12\rrbracket$ bivariate bicycle (BB) codes~\cite{bravyi2024high}.
The circuit setup is the same as Ref.~\cite{blue2025machine}, which follows the circuit designs from Ref.~\cite{bravyi2024high} (up to some small differences in the encoding round) to measure the check operators for $d$ rounds and the logical operators at the end. The PCM $H$, logical operator matrix $L$ and prior probabilities $\bm{p}$ for physical error events are generated using Stim~\cite{gidney2021stim}. The syndrome inputs include measurements from both $X$-type and $Z$-type check operators (i.e., those composed of only Pauli-$X$ or Pauli-$Z$ operators), and the logical error outputs correspond to only one type of logical operator, e.g., logical-$Z$ errors obtained by measuring the logical-$X$ operators~\cite{blue2025machine}. The neural networks are trained at a fixed physical error rate for the depolarization channel (Eq.~\ref{eq:depolar}), i.e., $p_\text{train}=0.006$ for both $\llbracket 72,12,6\rrbracket$ and $\llbracket 144,12,12\rrbracket$ BB codes, and are then used to decode across a range of other physical error rates without finetuning.

Meanwhile, as a baseline, we apply BP-OSD using CUDA-Q~\cite{NVIDIA_Corporation_CUDA-QX} on the same type of GPU as the neural decoders; after tuning the parameters, it displays faster decoding time and slightly better LER compared to the BP-OSD data from Ref.~\cite{blue2025machine}. Two protocols for BP-OSD have been considered: one uses the syndrome information from both $X$ and $Z$ check operators (BP-OSD(XZ)), and the other only uses $X$ check operator syndromes (BP-OSD(X)). 
BP-OSD(X) is more accurate at larger $p$ whereas BP-OSD(XZ) is more accurate at smaller $p$. 

Fig.~\ref{fig:circlevel} shows the total LER (i.e., not per round) and latency results of decoding $d$ rounds of syndrome measurements for the two BB codes.
For $\llbracket 72,12,6\rrbracket$ code, we observe that the masked DF decoder with $T=n_l=k$ has a lower LER than both the AR decoder and BP-OSD across the range of physical error rates $p$ displayed in Fig.~\pref{fig:circlevel}{a}; at small $p$, the accuracy of the DF decoder is approached by BP-OSD(XZ). 
Meanwhile, as shown in Fig.~\pref{fig:circlevel}{c}, the DF decoder and BP-OSD have similar decoding time, though the decoding time of BP-OSD increases with $p$, and the average latency transitions from faster to slower than the DF decoder as $p$ increases. 
For $\llbracket 144,12,12\rrbracket$ code, as shown in Fig.~\pref{fig:circlevel}{e}, the masked DF decoder is still among the ones giving the lowest LER, while the AR decoder does not maintain the advantage at $p=0.001$. 
However, the DF decoder shows faster average and worst-case decoding time than BP-OSD in this larger-scale code, as seen from Fig.~\pref{fig:circlevel}{g}. 

For the masked DF decoder, the total time step $T$ specifies the rate at which both the forward and reverse Markov chains approach their stationary distributions.  
One can use different values of $T$ for training and inference. Here, we investigate how decreasing $T$ during inference affects the speed and quality of the decoding. As $T$ gets smaller, more elements from the logical error prediction are unmasked per step; in the limit of $T=1$, all elements of $\bm{l}_0$ are unmasked at once. We observe that for $\llbracket 72,12,6\rrbracket$, there is very weak dependence of the LERs on $T$ (Fig.~\pref{fig:circlevel}{b}), while for $\llbracket 144,12,12\rrbracket$, LERs degrade slightly when reducing $T$ (Fig.~\pref{fig:circlevel}{f}). The decoding time scales approximately linearly with $T$ (Fig.~\pref{fig:circlevel}{d} and~\pref{fig:circlevel}{h}), making smaller $T$ non-trivially more efficient. 

Given that inference at $T=1$ still has reasonable accuracy, this motivates actually training at $T=1$ where the probability outputs are trained to predict $\bm{l}$ directly, reducing this framework to logistic regression (henceforth, we refer to the masked DF trained with $T=1$ as LR decoder). We observe that, for $\llbracket 72,12,6\rrbracket$, the LERs actually are improved marginally (Fig.~\pref{fig:circlevel}{a}), while for $\llbracket 144,12,12\rrbracket$, at small $p$ the LERs are worse than training at larger $T$ but still close in accuracy to the AR decoder (Fig.~\pref{fig:circlevel}{e}). 
It remains an open question whether these differences arise from model capacity limits, training-induced local minimum, or other factors. 

\subsection{Attention matrices from trained masked DF}
\label{sec:att_mask_compare}

\begin{figure*}
\centering
\includegraphics[width=\textwidth]{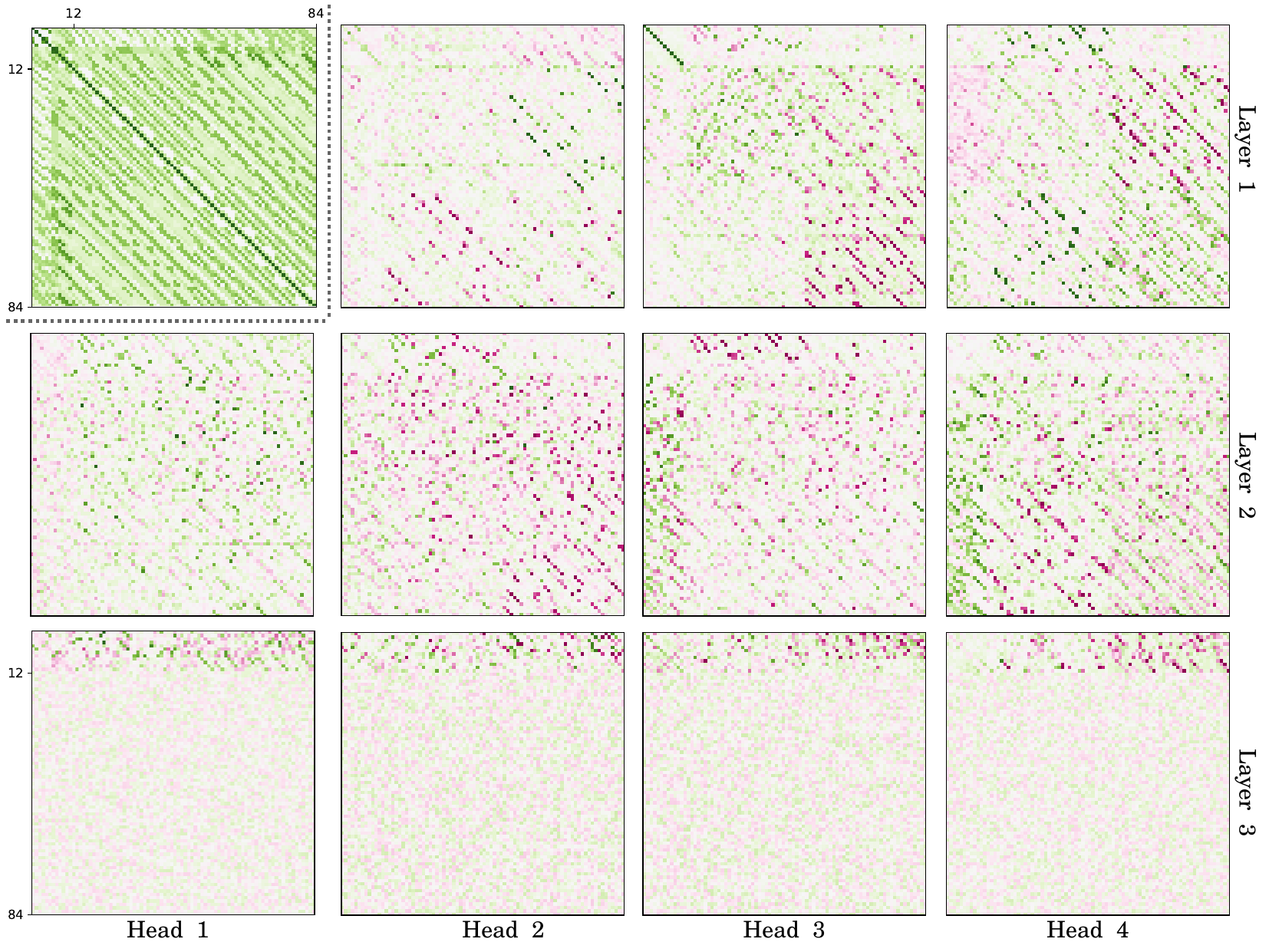}
\caption{Attention matrices from the multi-head factored-attention (MHFA) blocks for $\llbracket72,12,6\rrbracket$ circuit-level noise decoding. The top left one is the weight matrix $J$ constructed from the quantum code (see Methods~\ref{sec:PCMweight}), shown here for comparison. These attention matrices are initialized randomly from a Gaussian distribution before training. At the last layer (layer 3), the rows corresponding to check operators (i.e., row index below $n_c=12$) are not included in the loss function for optimization and stay random. Green/pink/white color denotes positive/negative/zero value; each plot has its own magnitude scale, hence color bar is not shown explicitly.}
\label{fig:attentionmat}
\end{figure*}

Within the neural network (see Fig.~\pref{fig:NN}{c}), the multi-head factored-attention (MHFA) blocks are the only places where hidden representations of elements in $\bm{l}$ or $\bm{s}$ exchange information. In Fig.~\ref{fig:attentionmat}, we display some trained attention matrices from different heads of the decoder layers, alongside the weight matrix $J$ constructed from the PCM $H$ and the logical matrix $L$ (see Sec.~\ref{sec:PCMweight}) for $\llbracket 72,12,6\rrbracket$ circuit-level noise decoding. This matrix $J$ defines the connectivity mediated by the shared physical errors between the syndrome bits and logical error bits, which is analogous to the Tanner graph structure. The Tanner graph is commonly used by some efficient structured decoders such as BP~\cite{mackay2004sparse}, as well as some recent well-performing neural decoders~\cite{maan2025machine,hu2025efficient,ataides2025neural}.
An interesting aspect from our work is that our decoder reveals this connectivity automatically via data-driven learning: from Fig.~\ref{fig:attentionmat}, we observe that the trained attention matrices display some features similar to the $J$ matrix. This is interesting since these attention matrices are initialized randomly, and during training, only paired syndrome–logical-error samples are fed into the neural networks, without any explicit knowledge of the quantum code or the error model.
These results suggest that efficient quantum decoding algorithms may come from those exploiting the quantum code structure.

\subsection{Decoding code-capacity error}
\begin{figure}
\centering
\hspace{-0.5cm}
\includegraphics[width=0.5\textwidth]{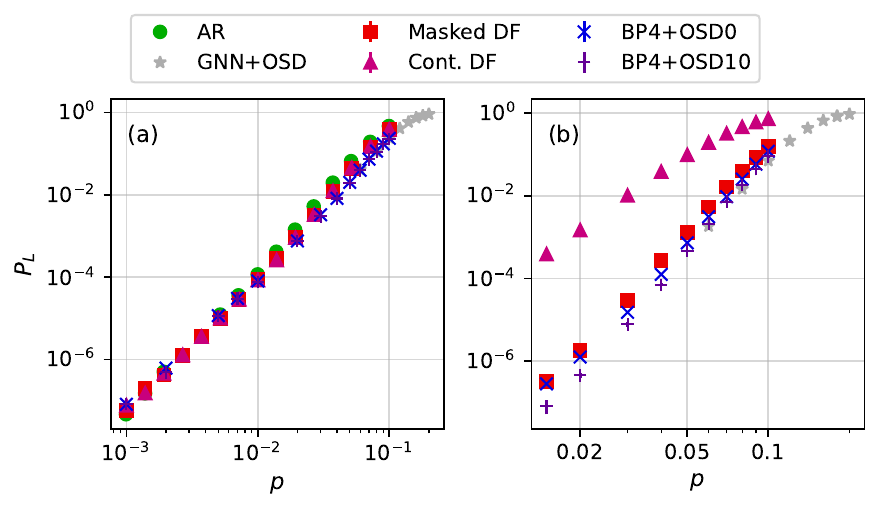}
\caption{Logical error rates of code capacity error decoded with continuous diffusion model (Cont. DF) and masked diffusion model (Masked DF) on (a) $\llbracket 72,12,6\rrbracket$ and (b) $\llbracket144,12,12\rrbracket$. Results from other decoders are included for comparison: AR decoder with data directly from Ref.~\cite{cao2025generative} without error bars; GNN+OSD decoder with data directly from Ref.~\cite{maan2025machine} without error bars; BP4+OSD0/10: quaternary BP with OSD of orders 0/10~\cite{gong2024graph,babar2015fifteen}.}
\label{fig:codecap}
\end{figure}

In this section, we now consider applying diffusion decoders not to circuit level noise but to code-capacity noise.  We continue using the same $\llbracket 72,12,6\rrbracket$ and $\llbracket144,12,12\rrbracket$ BB codes~\cite{bravyi2024high}. Here, both continuous DF and masked DF decoders are considered. The LERs from these DF decoders are compared with those from other decoders, including BP-OSD~\cite{panteleev2021degenerate}, AR decoder~\cite{cao2025generative}, and graph neural network (GNN) decoder~\cite{maan2025machine}. 

The input to the neural network contains both $X$ and $Z$-type syndromes, and the output is the prediction of logical errors for both the logical $X$ and $Z$ operators. The neural decoders are trained at a fixed physical error rate: $p_\text{train}=0.12$ for $\llbracket 72,12,6\rrbracket$ and $p_\text{train}=0.06$ for $\llbracket 144,12,12\rrbracket$, and are used to predict logical errors at other physical error rates $p$ without finetuning. 

Fig.~\pref{fig:codecap}{a} shows that, for $\llbracket 72,12,6\rrbracket$ BB code, all the decoders achieve quantitatively close LER across a wide range of physical error rates from $10^{-3}$ to $2\cdot10^{-2}$. One explanation could be that, within this range of $p$, all these decoders are already expressive enough to reach low enough logical error rates. However, when applying them to larger-scale quantum code (for example, $\llbracket 144,12,12\rrbracket$ BB code in Fig.~\pref{fig:codecap}{b}), because of the much larger sample space of $(\bm{e},\bm{s},\bm{l})$, the performance of data-driven neural decoders degrades given the limited number of training samples and the limited model capacity. BP-OSD performs the best, as the degradation of BP-OSD usually comes from the loops in their message-passing paths and multiple degenerate physical errors for the same syndrome~\cite{panteleev2021degenerate,babar2015fifteen}; and these limitations are not too severe for small- to medium-scale qLDPC codes. Nevertheless, from Fig.~\ref{fig:codecap}, we observe that masked DF performs better than continuous DF for the larger-scale $\llbracket 144,12,12\rrbracket$ BB code, and also shows improved LER over the smaller-scale $\llbracket 72,12,6\rrbracket$ BB code at the same $p$, suggesting better scalability from masked DF than the continuous DF.
In addition, the diffusion steps (and thus the decoding time) for continuous DF and masked DF can be different: while the former typically requires hundreds, the latter at most requires $T=n_l$, where one element from $\bm{l}$ is unmasked at each time step.

\section{Discussion}
In this work, we have demonstrated that diffusion (DF) models can serve as efficient neural decoders that infer logical errors from syndrome measurement outcomes for quantum error correction. Specifically, we investigate both continuous diffusion models and discrete masked diffusion models, applying them to bivariate bicycle (BB) quantum LDPC codes under both code-capacity and circuit-level noise. We evaluate the performance using two primary metrics: logical error rate (LER) and decoding time. 
Firstly, for circuit-level noise decoding, the masked DF decoder demonstrated improved LERs and decoding time over the autoregressive (AR) neural decoder. 
Compared to BP-OSD, the masked DF decoder consistently shows smaller LER (except for at $p=0.001$ on our larger code where it is effectively tied).  The masked DF-decoder is always much faster than the worst-case BP-OSD time.
The average speeds of the masked DF decoder are also faster for the larger code as well as the smaller code at large $p$.  For the smaller code at small $p$, there is a trade-off where BP-OSD(X) becomes faster while simultaneously becoming much less accurate than masked-DF. 
LR has slightly different trade-offs.  It is always faster (in expectation and for the worst case) than BP-OSD but becomes slightly less accurate at small $p$ than BP-OSD on the larger code. 
Interestingly, in our case, if one chooses the DF masked diffusion at $T=n_l/2$, one can get significant gains in speed at the expense of small to no degradation in LER. 

Next, for code-capacity noise decoding, we apply both continuous and masked DF decoders and observe reasonably low LERs, with the masked DF decoder exhibiting better scalability on larger codes. Here we also compared the diffusion decoders with other state-of-the-art qLDPC decoders: BP-OSD, AR and GNN neural decoders: for $\llbracket72,12,6\rrbracket$ code, all of them give quantitatively close LERs; for $\llbracket144,12,12\rrbracket$ code, masked DF decoder is shown to be only slightly worse than the best BP-OSD. 

The neural network architecture from this work is relatively simple, in particular for the masked DF decoder, where we replace the standard attention mechanism with factored-attention, so as to make the above improvements with significantly less parameters than the previous AR decoder. Interestingly, as the attention matrices from the factored-attention can be interpreted as the paths for exchanging messages among check operators and logical error bits during decoding, we observed that these paths capture some patterns from the quantum codes by learning from the error-syndrome data, informing a basic feature for constructing an efficient decoder. 

Nevertheless, several factors still need to be addressed for wider applications of diffusion decoders in future works.

\textit{Training time}: In this work, though we have adopted various strategies (e.g., multistage training) to speed up the training and reach convergence (see Methods.~\ref{sec:multi-stage}), in practice the training time is a bottleneck especially on larger-size codes and still requires further improvements. 
We expect that a well-designed neural network~\cite{maan2025machine,hu2025efficient,ataides2025neural}, careful tuning of hyperparameters, and training protocols such as transfer learning from neural decoders on smaller codes~\cite{maan2025machine}
can reduce the training time significantly.

\textit{Scalability}: Limited by training time, we have mainly considered the $\llbracket72,12,6\rrbracket$ and $\llbracket144,12,12\rrbracket$ BB codes in this work. However, it is worth investigating how well the diffusion decoders can be generalized to other larger-scale qLPDC codes, especially after incorporating various recent heuristic strategies (e.g., classifier-free guidance, adaptive layer norm~\cite{ho2022classifier,peebles2023scalable}) that were developed to improve diffusion models.

\textit{Other generative models}: Whether generative models other than auto-regressive and diffusion models, for example, flow-based models~\cite {lipman2022flow,ho2019flow}, would also exhibit promising performance in QEC decoding is an interesting direction to explore in the future.

\section{Methods}
\subsection{Notation}
Throughout this work, bold lowercase letters (e.g., $\bm{l}$) denote vectors; a subscript $t$ (e.g., $\bm{l}_t$) denotes the vector at time step $t$; and a subscript range $a:b$ denotes a sequence of vectors (e.g., $\bm{l}_{1:T}=\{\bm{l}_1,\bm{l}_2,\cdots,\bm{l}_T\}$, $d\bm{l}_{1:T}=d\bm{l}_1d\bm{l}_2\cdots d\bm{l}_T$). Individual vector elements use non-bold font with an additional index (e.g., $l_{i}$ is the $i$-th element of $\bm{l}$, $l_{t,i}$ is the $i$-th element of $\bm{l}_t$).

\subsection{Diffusion model loss function}
Training the diffusion decoder requires the minimization of the Kullback–Leibler (KL) divergence between $q_{\bm{\theta}}(\bm{l}|\bm{s})$ and $p(\bm{l}|\bm{s})$:
\begin{equation}
\label{eq:Loss0}
\mathcal{D}({\bm{\theta}})=\sum_{\bm{s}\in\{0,1\}^{n_s}}p(\bm{s})D_{\text{KL}}\left(p(\bm{l}|\bm{s})||q_{\bm{\theta}}(\bm{l}|\bm{s})\right)
\end{equation}
where $\bm{\theta}$ is the set of learnable parameters. However, a direct evaluation of $\mathcal{D}({\bm{\theta}})$ is not possible for diffusion models. An alternative and tractable loss function $\mathcal{L}(\bm{\theta})$ will be derived as an upper bound for $\mathcal{D}({\bm{\theta}})$. To this end, we factorize $q_{\bm{\theta}}(\bm{l}|\bm{s})$ as follows:
\begin{equation}
\label{eq:defMarkov}
q_{\bm{\theta}}(\bm{l}|\bm{s})=\sum_{\bm{l}_{1:T}}q_{\bm{\theta}}(\bm{l}_{0:T}|\bm{s})
=\sum_{\bm{l}_{1:T}} q(\bm{l}_T)\prod_{t=1}^{T} q_{\bm{\theta}}(\bm{l}_{t-1}|\bm{l}_t,\bm{s})
\end{equation}
where $\bm{l}\equiv\bm{l}_0$. It defines a Markov chain, with $q(\bm{l}_T)$ as the initial (prior) probability, and $q_{\bm{\theta}}(\bm{l}_{t-1}|\bm{l}_t,\bm{s})$ as the transition probability. We can contrast it with the AR decoders, where $q_{\bm{\theta}}(\bm{l}|\bm{s})$ is factorized as 
\begin{equation}
q_{\bm{\theta}}(\bm{l}|\bm{s})=q_{\bm{\theta}}(l_1|\bm{s})q_{\bm{\theta}}(l_2|l_1,\bm{s})\cdots q_{\bm{\theta}}(l_{n_l}|l_{1:n_l-1},\bm{s})
\end{equation}
and the elements of $\hat{\bm{l}}$ are generated sequentially via
\begin{equation}
\hat{l}_j=\arg\max_{l_j} q_{\bm{\theta}}(l_j|\hat{l}_{1:j-1},\bm{s})    
\end{equation}

Given Eq.~\eqref{eq:defMarkov}, the upper bound for $\mathcal{D}({\bm{\theta}})$ is shown to be (see Appendix~\ref{sec:theoryDF} for derivations)
\begin{equation}
\label{eq:lossTDKL}
\mathcal{L}(\bm{\theta})=\sum_{t=1}^T\mathbb{E}_{\bm{s},\bm{l},\bm{l}_t}D_{\text{KL}}(p(\bm{l}_{t-1}|\bm{l}_{t},\bm{l}_0,\bm{s})||q_{\bm{\theta}}(\bm{l}_{t-1}|\bm{l}_{t},\bm{s}))
\end{equation}
where $p(\bm{l}_{t-1}|\bm{l}_t,\bm{l}_0,\bm{s})$ is the ground-truth transition probability defined by the forward diffusion process. 

In the continuous diffusion model, the forward process is to add random Gaussian noise to $\bm{l}_t$, and $p(\bm{l}_{t-1}|\bm{l}_t,\bm{l}_0,\bm{s})$ is a Gaussian distribution with closed-form expressions for its mean and variance. Then $q_{\bm{\theta}}(\bm{l}_{t-1}|\bm{l}_t,\bm{s})$ can be defined as a Gaussian distribution with its mean value $\bm{\mu}_{\bm{\theta}}(\bm{l}_t,\bm{s},t)$ predicted by a neural network as follows: 
\begin{equation}
\label{eq:denoising}
\bm{\mu}_{\bm{\theta}}(\bm{l}_t,\bm{s},t)=\frac{1}{\sqrt{\alpha_t}}\left(\bm{l}_t-\frac{\beta_t}{\sqrt{1-\bar{\alpha}_t}}\bm{\epsilon}_{\bm{\theta}}(\bm{l}_{t},\bm{s},t)\right),   
\end{equation}
where $\beta_t$, $\alpha_t$ and $\bar{\alpha}_t$ are constants determined in the forward process, and $\bm{\epsilon}_{\bm{\theta}}(\bm{l}_{t},\bm{s},t)\in\mathbb{R}^{n_l}$ comes from the neural network. The training is reduced to the minimization of the distance between the mean values of $p(\bm{l}_{t-1}|\bm{l}_t,\bm{l}_0,\bm{s})$ and $q_{\bm{\theta}}(\bm{l}_{t-1}|\bm{l}_t,\bm{s})$ (see Appendix~\ref{sec:theoryDF}). 

In the masked diffusion models, the forward random process masks each element of $\bm{l}_t$ independently with probability $\beta_t$,\footnote{$\beta_t$ in masked diffusion is different from that in continuous diffusion.} and any element once masked remains masked for all subsequent time steps. Define $\beta_t=1/(T-t+1)$, such that at step $t$, $n_l\cdot t/T$ elements are masked in expectation, and one can directly sample $\bm{l}_t$ from $\bm{l}_0$ with masking probability $t/T$. After $T$ steps, all elements are guaranteed to be masked. 

Similar to the continuous diffusion models, the ground-truth transition probability $p(\bm{l}_{t-1}|\bm{l}_t,\bm{l}_0,\bm{s})$ has a closed-form expression as~\cite{austin2021structured,nie2025large,kim2025train,shi2024simplified} 
\begin{equation}
\label{eq:reverseProb}
p(\bm{l}_{t-1}|\bm{l}_t,\bm{l}_0,\bm{s})=\prod_{k=1}^{n_l}p(l_{t-1,k}|\bm{l}_{t},\bm{l}_0,\bm{s})  
\end{equation}
\begin{equation}
p(l_{t-1,k}|\bm{l}_{t},\bm{l}_{0},\bm{s})=
\begin{cases}
1, & l_{t,k}=l_{0,k},\;l_{t-1,k}=l_{t,k}\\
\frac{t-1}{t}, & l_{t,k}=*,\;l_{t-1,k}=*\\
\frac{1}{t}, & l_{t,k}=*,\;l_{t-1,k}=l_{0,k}\\
0, & \text{otherwise}
\end{cases}
\end{equation}
where `$*$' denotes being masked. By defining the variational probability in a similar form: 
\begin{equation}
\label{eq:variRevers_old}
q_{\bm{\theta}}(\bm{l}_{t-1}|\bm{l}_{t},\bm{s})=\prod_{k=1}^{n_l}q_{\bm{\theta}}(l_{t-1,k}|\bm{l}_{t},\bm{s})
\end{equation}
\begin{equation}
q_{\bm{\theta}}(l_{t-1,k}|\bm{l}_{t},\bm{s})
=\begin{cases}
1, & l_{t,k}\neq *,\;l_{t-1,k}=l_{t,k}\\
\frac{t-1}{t}, & l_{t,k}=*,\;l_{t-1,k}=* \\
\frac{1}{t}q_{\bm{\theta}}(l_{0,k}|\bm{l}_{t},\bm{s}), & l_{t,k}=*,\;l_{t-1,k}=l_{0,k}\\
0, & \text{otherwise}
\end{cases}
\end{equation}
the loss function Eq.~\eqref{eq:lossTDKL} is reduced to a simpler form
\begin{equation}
\mathcal{L}(\bm{\theta})=\mathbb{E}_{\bm{s},\bm{l}}\sum_{t=1}^{T}\frac1t\mathbb{E}_{\bm{l}_{t}\sim p(\bm{l}_{t}|\bm{l},\bm{s})}\sum_{\{k|l_{t,k}=*\}}\left[-\log q_{\bm{\theta}}(l_{0,k}|\bm{l}_{t},\bm{s})\right] 
\end{equation}
where the neural network directly predicts the probabilities for the original values of $\bm{l}_0$, analogous to the implicit diffusion model~\cite{song2020denoising}.

\subsection{Neural networks}
\label{sec:NNdetail}

\begin{figure*}
\centering
\includegraphics[width=\textwidth]{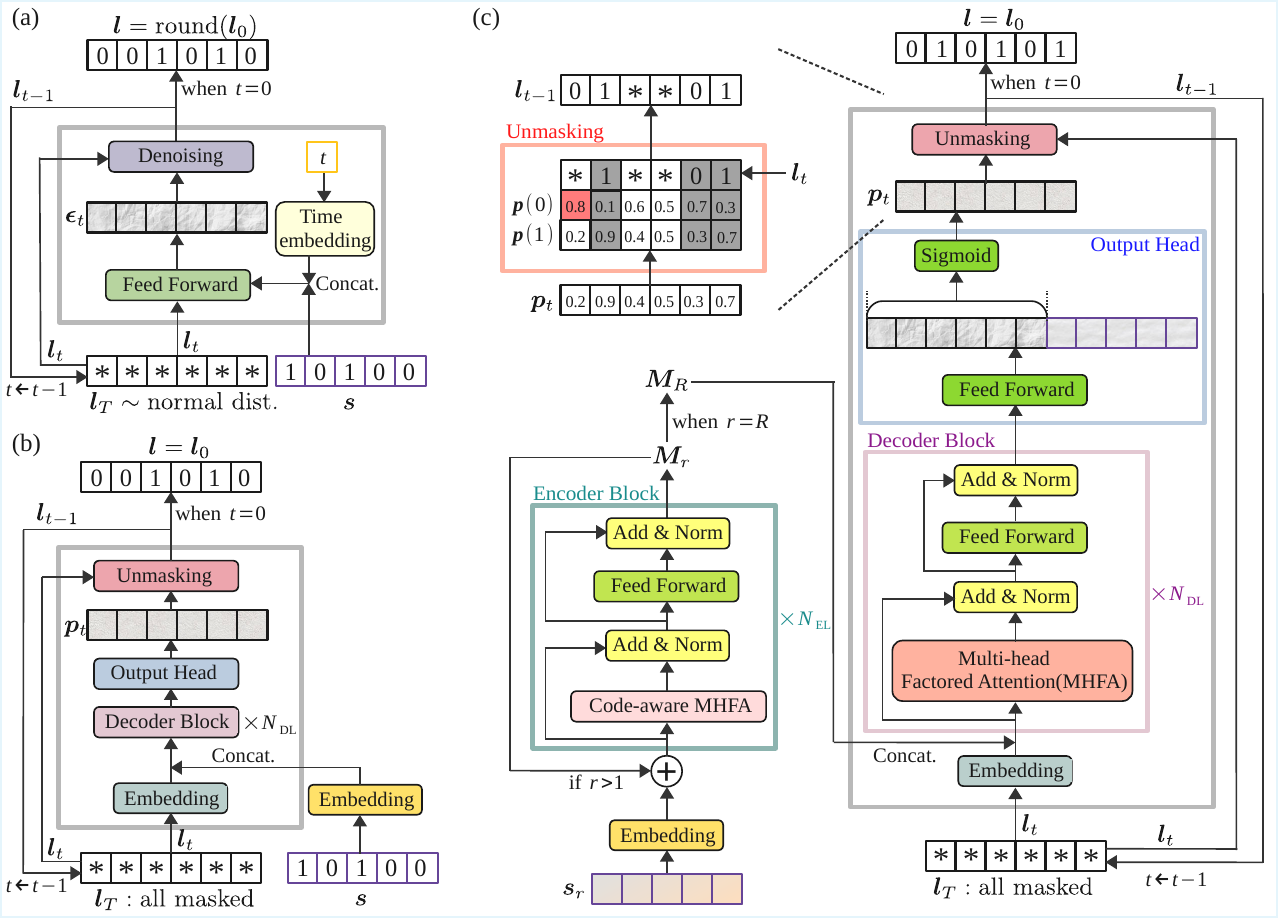}
\caption{Neural network architectures of (a) continuous diffusion model for code capacity error, (b) masked diffusion model for code capacity error, and (c) masked diffusion model for circuit-level error. The unmasking and decoder blocks in (b) are the same as and detailed in (c). Compared to (b), additional encoder blocks are inserted in (c) to process syndromes from each round of measurement.}
\label{fig:NN}
\end{figure*}

Here we provide more details on the neural network architectures of the DF decoders (see Fig.~\ref{fig:NN}). 

For continuous DF decoders on code-capacity noise decoding, the neural network takes $\bm{l}_t$, $\bm{s}$ and time step $t$ as the inputs. First, $t$ is embedded into a vector of dimension $d_t$: 
\begin{equation}
\begin{split}
\text{emb}(t)=&\bigl(\sin(\omega_0t), \sin(\omega_1t),\cdots,\sin(\omega_{(d_t-2)/2}t), \\
& \cos(\omega_0t), \cos(\omega_1t),\cdots,\cos(\omega_{(d_t-2)/2}t) \bigr)
\end{split}
\end{equation}
with $\omega_k=\frac{1}{1000^{2k/d_t}}$. Next, $\text{emb}(t)$ and $\bm{s}$ are concatenated to form a vector $\bm{y}\in\mathbb{R}^{d_t+n_s}$ as the condition input:
\begin{equation}
\bm{y} = \left( \text{emb}(t), \bm{s}-0.5\right)    
\end{equation}
where we shift $\bm{s}$ by a constant 0.5 to center it around zero. Then, to output $\bm{\epsilon}_{\bm{\theta}}(\bm{l}_t,\bm{s},t)$, we let $\bm{l}_t$ and $\bm{y}$ go through the following feed-forward layers:
\begin{equation}
\begin{split}
\bm{x}&\gets\left( \bm{l}_t-0.5, \bm{y} \right) \\
\bm{x}&\gets\text{gelu}\left(\bm{x}W_1+\bm{b_1}\right) \\
\bm{x}&\gets\text{gelu}\left(\bm{x}W_2+\bm{b_2}\right) \\
\bm{x}&\gets\left( \bm{x}, \bm{y} \right)\\
\bm{x}&\gets\text{gelu}\left(\bm{x}W_3+\bm{b_3}\right) \\
\bm{x}&\gets\text{gelu}\left(\bm{x}W_4+\bm{b_4}\right) \\
\bm{\epsilon}_{\bm{\theta}}(\bm{l}_t,\bm{s},t)&\gets \bm{x}W_5+\bm{b_5}
\end{split}    
\end{equation}
where $W_1\in\mathbb{R}^{(d_t+n_s)\times d_f}$, $\bm{b}_1\in\mathbb{R}^{d_f}$, $W_2\in\mathbb{R}^{d_f\times d_f}$, $\bm{b}_2\in\mathbb{R}^{d_f}$, $W_3\in\mathbb{R}^{(d_f+d_t+n_s)\times d_f}$, $\bm{b}_3\in\mathbb{R}^{d_f}$, $W_4\in\mathbb{R}^{d_f\times d_f}$, $\bm{b}_4\in\mathbb{R}^{d_f}$, 
$W_5\in\mathbb{R}^{d_f\times n_l}$, $\bm{b}_5\in\mathbb{R}^{n_l}$ are the trainable parameters in $\bm{\theta}$, and $d_f$ is the feedforward dimension in the hidden layers.

For the forward diffusion process, we use the linear scheduling for coefficient $\beta_t=\frac{0.1+19.9t/T}{T}$ with $T$ sufficiently large, e.g., $T=200$.

Since we will round $\bm{l}_0$ to integers, the last few denoising steps are less important, hence, at the training stage, we sample $t$ with probability proportional to the coefficient for $\bm{\epsilon}_{\bm{\theta}}(\bm{l}_{t},\bm{s},t)$ in Eq.~\eqref{eq:denoising}: $p(t)\propto\frac{\beta_t^2}{1-\bar{\alpha}_t}$, for which $t$ of larger values are more likely to be sampled. Following the standard diffusion model training algorithm~\cite{ho2020denoising}, we sample $t\sim p(t)$, $\bm{\epsilon}\sim\mathcal{N}(0,1)$, and $(\bm{s},\bm{l})$ from the error model; this gives a sample of $\bm{l}_t=\sqrt{\bar{\alpha}_t}\bm{l}+\sqrt{1-\bar{\alpha}_t}\bm{\epsilon}$. Note that we have shifted $\bm{l}$ from $\{0,1\}^{n_l}$ to $\{-0.5,0.5\}^{n_l}$ so as to keep the mean values of the variables around 0.

For the masked DF decoders on code-capacity noise decoding, we use the transformer architecture (see Fig.~\pref{fig:NN}{b,c}). First, given $\bm{l}_t$ and $\bm{s}$, we embed each of their elements into a vector of dimension $d_m$:
\begin{equation}
E^l(\bm{l}_t)=\left(\text{emb}^l(l_{t,1}),\text{emb}^l(l_{t,2}),\cdots,\text{emb}^l(l_{t,n_l}) \right)    
\end{equation}
\begin{equation}
\label{eq:emb_syn}
E^s(\bm{s})=\left(\text{emb}^s(s_1),\text{emb}^s(s_2),\cdots,\text{emb}^s(s_{n_s}) \right)    
\end{equation}
Notice that the masked entry `$*$' in $\bm{l}_t$ is also embedded like the 0 and 1 entries. Next, we concatenate $E^l(\bm{l}_t)\in\mathbb{R}^{d_m\times n_l}$ and $E^s(\bm{s})\in\mathbb{R}^{d_m\times n_s}$ into $\bm{x}\in\mathbb{R}^{d_m\times (n_l+n_s)}$:
\begin{equation}
\label{eq:concatenate}
\bm{x} = \left(E^l(\bm{l}_t), E^s(\bm{s})\right)
\end{equation}
and let $\bm{x}$ go through $N_{\text{DL}}$ layers of decoder blocks. Each decoder block gets input $\bm{x}\in\mathbb{R}^{d_m\times (n_l+n_s)}$ and outputs another $\bm{x}\in\mathbb{R}^{d_m\times (n_l+n_s)}$. Inside each decoder block, $\bm{x}$ is divided into $n_h$ attention heads after a linear map with $\{V\in\mathbb{R}^{d_m\times d_m}, \bm{b}_v\in\mathbb{R}^{d_m}\}$:
\begin{equation}
\label{eq:FA_p1}
\begin{split}
\bm{x}&\gets V\bm{x}+\bm{b}_v, \\
\bm{x}&=
\begin{pmatrix}
\bm{x}^{[1]} \\ \bm{x}^{[2]} \\ \vdots \\ \bm{x}^{[n_h]}    
\end{pmatrix}.
\end{split}
\end{equation}
For each $\bm{x}^{[h]}\in\mathbb{R}^{d_h\times(n_l+n_s)}$ (where $d_h=d_m/n_h$), an attention matrix $A^{[h]}$ is multiplied to it so that the hidden representations for each element of $\bm{l}_t$ or $\bm{s}$ can exchange messages from each other. 
In standard self-attention mechanism, $A^{[h]}$ come from the matrix product of query and key matrices, which are themselves dependent on the input $\bm{x}$; however, here we directly assign $A^{[h]}\in\mathbb{R}^{(n_l+n_s)\times(n_l+n_s)}$ as trainable parameters.
This so-called factored-attention (FA) mechanism has been widely used in neural network quantum states~\cite{viteritti2023transformer,viteritti2025quantum,rende2025foundation,viteritti2025transformer}. Lastly, the output from the multi-head factored attention is given by
\begin{equation}
\label{eq:FA_p2}
\bm{x}\gets \bm{b}_u+U    
\begin{pmatrix}
\bm{x}^{[1]}A^{[1]} \\ \bm{x}^{[2]}A^{[2]} \\ \vdots \\ \bm{x}^{[n_h]}A^{[n_h]}    
\end{pmatrix}.
\end{equation}
where $U\in\mathbb{R}^{d_m\times d_m}$, $\bm{b}_u\in\mathbb{R}^{d_m}$ are trainable parameters. Note that the vectors $\bm{b}_u$, $\bm{b}_v$ are broadcast to each column when added to a matrix. Combining Eq.~\eqref{eq:FA_p1} and Eq.~\eqref{eq:FA_p2}, we have $\bm{x}\gets\text{MHFA}(\bm{x})$, and computations inside a decoder block can be summarized as:
\begin{equation}
\begin{split}
\bm{x}&\gets\text{LayerNorm}\left(\bm{x}+\text{MHFA}(\bm{x})\right) \\
\bm{x}&\gets\text{LayerNorm}\left(\bm{x}+\text{FF}(\bm{x})\right)
\end{split}
\end{equation}
where the feed-forward (FF) layer performs the following computation:
\begin{equation}
\text{FF}(\bm{x})= \bm{b}_2+W_2\;\text{gelu}\left(\bm{b}_1+W_1\bm{x}\right)   
\end{equation}
where $W_1\in\mathbb{R}^{d_f\times d_m}$, $\bm{b}_1\in\mathbb{R}^{d_f}$, $W_2\in\mathbb{R}^{d_m\times d_f}$, $\bm{b}_2\in\mathbb{R}^{d_m}$ are the trainable parameters.

Finally, the output head consists of one feed-forward layer, to reduce the dimension of the hidden vectors for $\bm{l}_t$ to 1:
\begin{equation}
\bm{p}_t\gets\text{sigmoid}\left( \bm{b}_0+W_0\bm{x}_{1:n_l}\right)  
\end{equation}
with $W_0\in\mathbb{R}^{1\times d_m}$, $\bm{b}_0\in\mathbb{R}$ as trainable parameters, such that we can use $\bm{p}_t$ as the probabilities for each masked element in $\bm{l}_t$ to take value 1, i.e., $q_{\bm{\theta}}(l_{t-1,k}=1|\bm{l}_t,\bm{s})=p_{t,k}$ and $q_{\bm{\theta}}(l_{t-1,k}=0|\bm{l}_t,\bm{s})=1-p_{t,k}$.

For the masked DF decoders on circuit-level noise decoding, we replace the embedding layer for syndromes with additional encoder blocks (see Fig.~\pref{fig:NN}{c}). Once we have processed the syndrome from all the $R$ rounds and obtained $M_R\in\mathbb{R}^{d_m\times n_c}$ --- $n_c$ is the number of check operators, equal to the $n_s$ from code-capacity noise model, but for circuit-level noise model, $n_c=n_s/(R+1)$ --- as the hidden representation for the syndrome $\bm{s}$, we obtain $\bm{x}$ by concatenation as in Eq.~\eqref{eq:concatenate}:
\begin{equation}
\bm{x}=\left( E^l(\bm{l}_t), M_R\right)    
\end{equation}
then the rest steps follow the same as the code-capacity noise decoding.

Now we explain the encoder block in detail. In the syndrome-measurement circuit from Ref.~\cite{blue2025machine}, there are two rounds noiseless measurements before and after the $R$ rounds noisy measurements on the check operators, respectively; the parity checking between successive rounds generates the syndrome $\bm{s}\in\{0,1\}^{n_s}$ with $n_s=n_c(R+1)$. Therefore, we divide $\bm{s}$ into $R+1$ blocks
\begin{equation}
\label{eq:syndroblock}
\bm{s}=\left(\bm{s}_0,\bm{s}_1,\cdots,\bm{s}_R\right),    
\end{equation}
and process one block at a time.  

To be specific, when it comes to $\bm{s}_r$, we first embed it as $E^s(\bm{s})\in\mathbb{R}^{d_m\times n_c}$ (Eq.~\eqref{eq:emb_syn}) and add it to the output $M_{r-1}$ from last round: 
\begin{equation}
\bm{y}=M_{r-1}+E^s(\bm{s}_r)    
\end{equation}
Note that for $r=0$, $\bm{y}=E^s(\bm{s}_0)$.

Next, we let $\bm{y}$ go through $N_{\text{EL}}$ encoder blocks to get $M_r$. The encoder block has the same structure as the decoder block, except that inside MHFA, each attention matrix from each encoder block (e.g., head $h$ from block $b$) $A^{[b,h]}$ is element-wise multiplied with a weight matrix $K^{[r]}\in\mathbb{R}^{(n_l+n_c)\times(n_l+n_c)}$ that depends on the round index $r$:
\begin{equation}
\label{eq:attentionMatMask}
\widetilde{A}^{[b,h,r]}_{jk}=A^{[b,h]}_{jk}K^{[r]}_{jk}    
\end{equation}
before applying to the input $\bm{x}$. $\{K^{[r]}\}_{r=0}^{R}$ are initialized with the PCM $H$ of the quantum code (see Sec.~\ref{sec:PCMweight} for more details), and are trainable parameters, hence we refer to this MHFA with weight matrices as code-aware MHFA~\cite{blue2025machine,choukroun2022error}. 

We have also leveraged JAX’s just-in-time compilation~\cite{jax2018github} to further accelerate the training and inference. Finally, in Tab.~\ref{tab:NN_hyper} we display the hyperparameters for the neural networks. It also shows that the DF decoders have much less parameters than the previous AR decoders.

\begin{table}[h]
\caption{\label{tab:NN_hyper}%
Hyperparameters for neural networks.}
\begin{ruledtabular}
\begin{tabular}{lcc}
Cont. DF (code-capacity) &
$\llbracket 72,12,6 \rrbracket$ &
$\llbracket 144,12,12 \rrbracket$ \\
\colrule
Diffusion steps ($T$) & 200 & 200\\
Feedforward dimension ($d_m$) & 2,048 & 2,048\\
Total parameter number & 13,557,784 & 13,852,696\\
\colrule
Masked DF (code-capacity) & & \\
\colrule
Decoder layers ($N_{\text{DL}}$) & 4 & 32 \\
Attention heads ($n_h$) & 8 & 8 \\
Model dimension ($d_m$) & 128 & 96 \\
Feedforward dimension ($d_f$) & 512 & 384\\
Total parameter number & 956,929 & 11,112,193\\
\colrule
AR decoder (code-capacity)~\cite{cao2025generative} & & \\
\colrule
Total parameter number & 16,289,364 & $\backslash$\\
\colrule
Masked DF (circuit-level) & & \\
\colrule
Encoder layers ($N_{\text{EL}}$) & 3 & 3 \\
Decoder layers ($N_{\text{DL}}$) & 3 & 3 \\
Attention heads ($n_h$) & 8 & 8 \\
Model dimension ($d_m$) & 256 & 512 \\
Feedforward dimension ($d_f$) & 512 & 1,024\\
Total parameter number & 2,705,217 & 10,820,225\\
\colrule
AR decoder (circuit-level)~\cite{blue2025machine} & & \\
\colrule
Total parameter number & $4.77\times10^6$ & $1.90\times10^7$
\end{tabular}
\end{ruledtabular}
\end{table}

\subsection{Weight matrix from circuit-level PCM}
\label{sec:PCMweight}
Consider a circuit-level PCM $H\in\mathbb{F}_2^{n_s\times n_e}$ for $R$ rounds of circuit-level measurements. We can divide $H$ into blocks as following
\begin{equation}
H=
\begin{pmatrix}
H^{[0]} \\ H^{[1]} \\ \vdots \\ H^{[R]}    
\end{pmatrix}
\end{equation}
Each block matrix $H^{[r]}\in\mathbb{F}_2^{n_c\times n_e}$ ($n_c$ is the number of check operators) generates the corresponding syndrome $\bm{s}_r=H^{[r]}\hat{\bm{e}}$ from  Eq.~\eqref{eq:syndroblock}. Notice that each element from $\bm{s}_r\in\mathbb{F}_2^{n_c}$ corresponds to a check operator, and $H^{[r]}_{ij}=1$ indicates that the error event $j$ affects the syndrome corresponding to check operator $i$ at $r$-th round. Let 
\begin{equation}
\widetilde{H}^{[r]}_{ij}=
\begin{cases}
1, &\text{if } \left(\sum_{r'=0}^rH^{[r']}_{ij}\right)>0 \\
0, &\text{otherwise}
\end{cases}
\end{equation}
then $\widetilde{H}^{[r]}_{ij}=1$ indicates that the error event $j$ affects the syndrome measurement corresponding to the check operator operator $i$ from 0-th to $r$-th round. $\widetilde{K}^{[r]}=\widetilde{H}^{[r]}\widetilde{H}^{[r]T}$ gives the correlation among the check operators, i.e., $\widetilde{K}^{[r]}_{jk}$ is equal to the total number of error events that affect check operator $j$ and $k$ simultaneously. 

Similarly, we can establish the correlation between the logical error and check operators by defining
\begin{equation}
\widetilde{L}^{[R]}=
\begin{pmatrix}
L\\ \widetilde{H}^{[R]}    
\end{pmatrix}
\end{equation}
and $J^{[R]}=\widetilde{L}^{[R]}\widetilde{L}^{[R]T}$. 

In our BERT-based neural network, we identify each column of the output matrix (e.g., $M_r$ from the encoder blocks) as a hidden representation for the corresponding input from a check operator or logical error. The attention blocks are the only places where these hidden representations interact with each other through the attention matrices $A$. Inspired by the BP decoders which pass messages between correlated check operators and error events, as well as the AR decoders which multiply the attention matrices with weight matrices constructed from the PCM~\cite{blue2025machine,choukroun2022error}, we use weight matrices $K^{[r]}$ to modify the attention matrices from the encoder blocks at each round $r$ (see Eq.~\eqref{eq:attentionMatMask}) and initialize it as $K^{[r]}_{jk}=\left(\widetilde{K}^{[r]}_{jk}\right)^{\frac18}$, where the eighth root is used heuristically to ensure that $K^{[r]}_{jk}$ has similar magnitude. 

On the other hand, the decoder blocks from our neural network do not have this weight matrix modification. In Sec.~\ref{sec:att_mask_compare}, we show that the attention matrices from the decoder blocks display the patterns of $J\equiv J^{[R]}$ after training from the syndrome-error data.

\subsection{Multi-stage training under circuit-level noise}
\label{sec:multi-stage}

Ref.~\cite{blue2025machine} proposes the following multi-stage training procedure. When building the dataset, in addition to all the detectors triggered by the faults, one also calculates the intermediate logical observables at the end of each syndrome extraction (SE) cycle. One would expect the neural network to give correct intermediate steps that correspond exactly to those logical observables. However, this is impossible at intermediate steps where noiseless SE results are not yet available. Moreover, with probability (on the order of) $p$, where $p$ is the physical error rate, the true logical observable will be different 
from what is noted down in the dataset: imagine the last CNOT gate in a noisy SE cycle, say it has control on a data qubit and target on an ancilla qubit to be measured in the $Z$ basis. One cannot differentiate the two fault mechanisms, $XX$ and $IX$, after this CNOT, without noiseless SE. This is because both faults have the same effect on the detectors, while triggering different logical observables. Nevertheless, intermediate results are still valuable resources for training under circuit-level noise. The multi-state training \cite{blue2025machine} makes use of them in the following way: one gradually shifts away from taking loss with intermediate predictions and logical observables at \emph{every} round, to taking loss only at the last (noiseless) SE round.

Formally, as illustrated in Sec.~\ref{sec:NNdetail} and Fig.~\pref{fig:NN}{c}, each time we process the syndrome from $r$-th round, we obtain the corresponding hidden representation $M_r$ for all the check operators. At the decoding stage, we send $M_R$ to the decoder blocks; however, within the multi-stage training, we can send the intermediate representation $M_r$ to decoder blocks to generate the probability $\bm{p}^{[r]}_t$ for intermediate logical error. Meanwhile, we get the target intermediate logical error $\bm{l}^{[r]}$ as follows: notice that from $\widetilde{H}^{[r]}$ we know which errors affect the check operators up to round $r$; we can build a projection operator $\Pi^{[r]}\in\mathbb{R}^{n_e}$ with
\begin{equation}
\Pi^{[r]}_{k}=
\begin{cases}
1, &\text{if } \sum_{j=1}^{n_c}\widetilde{H}^{[r]}_{jk}>0 \\
0, &\text{otherwise}
\end{cases}
\end{equation}
and get the intermediate physical error $\bm{e}^{[r]}$ with $e^{[r]}_k=\Pi^{[r]}_{k}e_k$ as well as  $\bm{l}^{[r]}=L\bm{e}^{[r]}$.
Finally, we substitute $\bm{p}^{[r]}_t$ and $\bm{l}^{[r]}$ into Eq.~\eqref{eq:lossMaskFinal} as the loss function for this intermediate decoding $\mathcal{L}^{[r]}(\bm{\theta})$. 

During the multi-stage training, we define a loss function 
\begin{equation}
\mathcal{L}(\bm{\theta}; R_1, R_2)=\sum_{r=R_1}^{R_2}\mathcal{L}^{[r]}(\bm{\theta})   
\end{equation}
for each stage. Typically, we set $(R_1,R_2)=(0,R)$ at the first stage, and gradually increase $R_1$ until $R_1=R$ at the final stage. 

In this work, we use AdamW (with first moment decay rate 0.9, second moment decay rate 0.999 and weight decay 0.0001) as the optimizer~\cite{loshchilov2017decoupled}, and the hyperparameters for training are listed in Tab.~\ref{tab:train_hyper}. They are all trained with a single NVIDIA A100 GPU.

\begin{table}[h]
\caption{\label{tab:train_hyper}%
Hyperparameters for training.}
\begin{ruledtabular}
\begin{tabular}{lcc}
Cont. DF (code-capacity) &
$\llbracket 72,12,6 \rrbracket$ &
$\llbracket 144,12,12 \rrbracket$ \\
\colrule
Batch size & 250 & 250\\
Learning rate\footnote{For code-capacity noise model, we initially set a learning rate, e.g., $10^{-3}$, then use cosine decay schedule to decrease it to a lower learning rate, e.g., $10^{-5}$, within $\approx 2000$ iterations.} & $10^{-3}\sim10^{-5}$ & $10^{-3}\sim10^{-5}$\\
Total training iterations & $8\times10^7$ & $8\times10^7$\\
Training time & $\approx$32 hours & $\approx$42 hours\\
\colrule
Masked DF (code-capacity) & & \\
\colrule
Batch size & 250 & 250\\
Learning rate & $10^{-4}\sim10^{-5}$ & $10^{-4}\sim10^{-5}$\\
Total training iterations & $1.6\times10^7$ & $8.8\times10^6$\\
Training time & $\approx$4 days & $\approx$4 days\\
\colrule
Masked DF (circuit-level) & & \\
\colrule
Batch size & 250 & 125\\
Learning rate\footnote{For circuit-level noise model, we initially set a smaller learning rate, e.g., $10^{-6}$, then use linear warm-up schedule to increase it to a larger learning rate, e.g., $10^{-4}$, within $\approx 2000$ iterations.} & $10^{-6}\sim10^{-4}$ & $10^{-6}\sim10^{-4}$\\
\colrule
Multi-stage training & & \\
$\{(R_1, R_2),\text{iterations}\}$ & & \\
1 & $(0,6),8\times10^4$ & $(0,6),1.6\times10^5$ \\
2 & $(1,6),8\times10^4$ & $(1,6),1.6\times10^5$ \\
3 & $(2,6),8\times10^4$ & $(2,6),1.6\times10^5$ \\
4 & $(3,6),8\times10^4$ & $(3,6),1.6\times10^5$ \\
5 & $(4,6),8\times10^4$ & $(4,6),1.6\times10^5$ \\
6 & $(5,6),8\times10^4$ & $(5,6),1.6\times10^5$ \\
7 & $(6,6),7.2\times10^5$ & $(6,6),1.6\times10^5$ \\
Change learning rate & $\backslash$ & $10^{-7}\sim10^{-5}$\\
8 &  & $(6,12),9.6\times10^4$ \\
9 &  & $(7,12),9.6\times10^4$ \\
10 &  & $(8,12),9.6\times10^4$ \\
11 &  & $(9,12),9.6\times10^4$ \\
12 &  & $(10,12),9.6\times10^4$ \\
13 &  & $(11,12),9.6\times10^4$ \\
14 &  & $(12,12),2.9\times10^5$ \\
\colrule
Training time & $\approx$32 hours & $\approx$10 days\\
\end{tabular}
\end{ruledtabular}
\end{table}

\subsection{BP-OSD}

In circuit-level noise decoding, we use the BP-OSD from CUDA-Q. For both the BP-OSD(X) and BP-OSD(XZ) settings, we use a maximum iteration of $1000$ for min-sum BP (scaling factor $1.0$), followed by OSD with combination sweep of order three.

In the code-capacity noise decoding, we use the quaternary BP (BP4)-OSD \cite{panteleev2021degenerate} implemented from Ref.~\cite{gong2024toward}. For a detailed explanation of quaternary BP \cite{fifteen-years}, see Ref.~\cite[Ch.~3]{gong2024graph}. For BP, the maximum iteration is set to $100$ and the min-sum scheduling is used. The OSD post-processing stage solves the $X$ and $Z$ syndrome equations separately, as in Ref.~\cite{panteleev2021degenerate}. However, one difference to Ref.~\cite{panteleev2021degenerate} is that the variable nodes are ranked according to their likelihood of being flipped \cite{roffe2020decoding} rather than their reliability when solving the syndrome equations. Another difference is that we use a scaling factor of $0.5$ in the min-sum BP check-node update, since this number performs better than the original factor $0.625$ proposed in Ref.~\cite{panteleev2021degenerate}.

\begin{acknowledgments}    
We thank J. Blue, J. Hahm and A. Senior for enlightening discussion. Z.L. and B.K.C. acknowledge support from the NSF Quantum Leap Challenge Institute for Hybrid Quantum Architectures and Networks (NSF Award No. 2016136). This research used the Delta advanced computing and data resource which is supported by the National Science Foundation (award OAC 2005572) and the State of Illinois, through allocation PHY240122 from the Advanced Cyberinfrastructure Coordination Ecosystem: Services \& Support (ACCESS) program, which is supported by National Science Foundation grants \#2138259, \#2138286, \#2138307, \#2137603, and \#2138296~\cite{citeACCESS}. Delta is a joint effort of the University of Illinois Urbana-Champaign and its National Center for Supercomputing Applications.
\end{acknowledgments}

\appendix
\section{Theory of diffusion decoder}
\label{sec:theoryDF}
This section presents the theoretical formulation of diffusion decoders, justifying their applicability to QEC decoding~\cite{cao2025generative,luo2022understanding,austin2021structured}.

The goal of a neural decoder is to model the data distribution $p(\bm{l}|\bm{s})$ with a variational probability $q_{\bm{\theta}}(\bm{l}|\bm{s})$, where $(\bm{l},\bm{s})$ are implicitly sampled from the prior probability $\bm{p}$ for physical error $\hat{\bm{e}}$.
To this end, a loss function is defined for optimization: the KL divergence between $q_{\bm{\theta}}(\bm{l}|\bm{s})$ and $p(\bm{l}|\bm{s})$
\begin{equation}
\mathcal{L}({\bm{\theta}})=\sum_{\bm{s}\in\{0,1\}^{n_s}}p(\bm{s})D_{\text{KL}}\left(p(\bm{l}|\bm{s})||q_{\bm{\theta}}(\bm{l}|\bm{s})\right)
\end{equation}
Note that 
\begin{equation}
D_{\text{KL}}\left(p(\bm{l}|\bm{s})||q_{\bm{\theta}}(\bm{l}|\bm{s})\right)=\sum_{\bm{l}\in\{0,1\}^{n_l}} p(\bm{l}|\bm{s}) \log \frac{p(\bm{l}|\bm{s})}{q_{\bm{\theta}}(\bm{l}|\bm{s})}   
\end{equation}
we rewrite the loss function as
\begin{equation}
\label{eq:lossLS}
\mathcal{L}({\bm{\theta}})=\sum_{\bm{s},\bm{l}}p(\bm{l},\bm{s})\log\frac{p(\bm{l}|\bm{s})}{q_{\bm{\theta}}(\bm{l}|\bm{s})}
\end{equation}
In practice, we estimate Eq.~\eqref{eq:lossLS} by sampling $(\bm{l},\bm{s})$ from $p(\bm{l},\bm{s})=\sum_{\bm{e}\in\{\bm{e}|H\bm{e}=\bm{s},L\bm{e}=\bm{l}\}}\Pr[\bm{e}]$, which yields
\begin{equation}
\mathcal{L}(\bm{\theta})=\mathbb{E}_{\bm{e}\sim\Pr[\bm{e}]}\{ -\log q_{\bm{\theta}}(\bm{l}(\bm{e})|\bm{s}(\bm{e})) \}   
\end{equation}
Note that terms independent of $\bm{\theta}$ can be added or removed without affecting optimization. For convenience, we turn it back to 
\begin{equation}
\label{eq:lossLogLS}
\mathcal{L}(\bm{\theta})=\mathbb{E}_{\bm{s},\bm{l}}\{-\log q_{\bm{\theta}}(\bm{l}|\bm{s})\}    
\end{equation}
without showing $\bm{e}$ explicitly. 
To facilitate the training of DF decoders, rather than optimizing the loss function Eq.~\eqref{eq:lossLogLS} directly, we substitute Eq.~\eqref{eq:defMarkov} into it and derive an upper bound.

Before proceeding, we give useful identities for the forward process:
\begin{equation}
\begin{split}
p(\bm{l}_{1:T}|\bm{l}_0,\bm{s})=&p(\bm{l}_1|\bm{l}_0,\bm{s})\prod_{t=2}^T p(\bm{l}_t|\bm{l}_{t-1}, \bm{l}_0,\bm{s})\\
=&p(\bm{l}_1|\bm{l}_0,\bm{s})\prod_{t=2}^T\frac{p(\bm{l}_{t-1}|\bm{l}_t,\bm{l}_0,\bm{s})p(\bm{l}_t|\bm{l}_0,\bm{s})}{p(\bm{l}_{t-1}|\bm{l}_0,\bm{s})}\\
=&p(\bm{l}_T|\bm{l}_0,\bm{s})\prod_{t=2}^T p(\bm{l}_{t-1}|\bm{l}_t,\bm{l}_0,\bm{s})
\end{split}
\end{equation}
The derivations that follow are standard for diffusion models~\cite{ho2020denoising,luo2022understanding}.
\begin{equation}
\label{eq:ELBO}
\begin{split}
&\mathcal{L}(\bm{\theta})=\mathbb{E}_{\bm{s},\bm{l}}\{-\log\sum_{\bm{l}_{1:T}}q_{\bm{\theta}}(\bm{l}_{0:T}|\bm{s})\}\\
&=\mathbb{E}_{\bm{s},\bm{l}} \Bigl\{-\log\sum_{\bm{l}_{1:T}}p(\bm{l}_{1:T}|\bm{l}_0,\bm{s})\frac{q_{\bm{\theta}}(\bm{l}_{0:T}|\bm{s})}{p(\bm{l}_{1:T}|\bm{l}_0,\bm{s})}\Bigl\}\\
&=\mathbb{E}_{\bm{s},\bm{l}} \Bigl\{-\log\mathbb{E}_{\bm{l}_{1:T}}\frac{q_{\bm{\theta}}(\bm{l}_{0:T}|\bm{s})}{p(\bm{l}_{1:T}|\bm{l}_0,\bm{s})}\Bigl\} \\
&\leq\mathbb{E}_{\bm{s},\bm{l}_{0:T}}\Bigl\{-\log\frac{q_{\bm{\theta}}(\bm{l}_{0:T}|\bm{s})}{p(\bm{l}_{1:T}|\bm{l}_0,\bm{s})}\Bigl\}\quad \text{(Jensen’s inequality)} \\
&=\mathbb{E}_{\bm{s},\bm{l}_{0:T}}\Bigl\{-\log\frac{q(\bm{l}_T)\prod_{t=1}^T q_{\bm{\theta}}(\bm{l}_{t-1}|\bm{l}_t,\bm{s})}{\prod_{t=1}^T p(\bm{l}_t|\bm{l}_{t-1},\bm{l}_0,\bm{s})}\Bigl\} \\
&=\mathbb{E}_{\bm{s},\bm{l}_{0:T}}\Bigl\{-\log\frac{q(\bm{l}_T)\prod_{t=1}^T q_{\bm{\theta}}(\bm{l}_{t-1}|\bm{l}_t,\bm{s})}{p(\bm{l}_T|\bm{l}_0,\bm{s})\prod_{t=2}^T p(\bm{l}_{t-1}|\bm{l}_t,\bm{l}_0,\bm{s})}\Bigl\}\\
&=\mathbb{E}_{\bm{s},\bm{l}_{0:T}}\Bigl\{-\log\frac{q(\bm{l}_T)}{p(\bm{l}_T|\bm{l}_0,\bm{s})}\Bigl\}+\mathbb{E}_{\bm{s},\bm{l}_{0:T}}\Bigl\{-\log q_{\bm{\theta}}(\bm{l}_0|\bm{l}_1,\bm{s})\Bigl\}\\
&+\sum_{t=2}^T\mathbb{E}_{\bm{s},\bm{l}_{0:T}}\Bigl\{-\log\frac{q_{\bm{\theta}}(\bm{l}_{t-1}|\bm{l}_t,\bm{s})}{p(\bm{l}_{t-1}|\bm{l}_t,\bm{l}_0,\bm{s})}\Bigl\}\\
&=\underbrace{\mathbb{E}_{\bm{s},\bm{l}_0,\bm{l}_T}\Bigl\{\log\frac{p(\bm{l}_T|\bm{l}_0,\bm{s})}{q(\bm{l}_T)}\Bigl\}}_{\mathcal{L}_T}+\underbrace{\mathbb{E}_{\bm{s},\bm{l}_0,\bm{l}_1}\Bigl\{-\log q_{\bm{\theta}}(\bm{l}_0|\bm{l}_1,\bm{s})\Bigl\}}_{\mathcal{L}_0(\bm{\theta})}\\
&+\sum_{t=2}^T\underbrace{\mathbb{E}_{\bm{s},\bm{l}_0,\bm{l}_{t-1},\bm{l}_t}\log\frac{p(\bm{l}_{t-1}|\bm{l}_t,\bm{l}_0,\bm{s})}{q_{\bm{\theta}}(\bm{l}_{t-1}|\bm{l}_t,\bm{s})}}_{\mathcal{L}_{t-1}(\bm{\theta})}
\end{split}
\end{equation}    
In the final expression of Eq.~\eqref{eq:ELBO}, $\mathcal{L}_T$ is the prior matching term that measures the discrepancy between the distribution of the fully noised state $\bm{l}_T$ and the simple prior distribution $p(\bm{l}_T)$; this term depends solely on the design of the diffusion protocol. The set of terms $\{\mathcal{L}_t\}_{t=0}^{T-1}$ are the denoising matching terms giving the KL divergence between the ground-truth denoising transition probability $p(\bm{l}_{t}|\bm{l}_{t+1},\bm{l}_0,\bm{s})$ and the variational denoising probability $q_{\bm{\theta}}(\bm{l}_t|\bm{l}_{t+1},\bm{s})$. To see this, we show that, for $\{\mathcal{L}_t\}_{t=1}^{T-1}$,
\begin{equation}
\begin{split}
&\mathcal{L}_t=\sum_{\bm{s},\bm{l}_0,\bm{l}_t,\bm{l}_{t+1}}p(\bm{s},\bm{l}_0,\bm{l}_t,\bm{l}_{t+1})\log\frac{p(\bm{l}_{t}|\bm{l}_{t+1},\bm{l}_0,\bm{s})}{q_{\bm{\theta}}(\bm{l}_{t}|\bm{l}_{t+1},\bm{s})} \\
&=\sum_{\bm{s},\bm{l}_0,\bm{l}_t,\bm{l}_{t+1}}p(\bm{s},\bm{l}_0,\bm{l}_{t+1})p(\bm{l}_t|\bm{l}_{t+1},\bm{s},\bm{l}_0)\log\frac{p(\bm{l}_{t}|\bm{l}_{t+1},\bm{l}_0,\bm{s})}{q_{\bm{\theta}}(\bm{l}_{t}|\bm{l}_{t+1},\bm{s})} \\
&=\sum_{\bm{s},\bm{l}_0,\bm{l}_{t+1}}p(\bm{s},\bm{l}_0,\bm{l}_{t+1})D_{\text{KL}}(p(\bm{l}_{t}|\bm{l}_{t+1},\bm{l}_0,\bm{s})||q_{\bm{\theta}}(\bm{l}_{t}|\bm{l}_{t+1},\bm{s}))
\end{split}
\end{equation}
and for $\mathcal{L}_0$, we can add terms with
\begin{equation}
\label{eq:reconstruct_prob_add}
p(\bm{l}|\bm{l}_1,\bm{l}_0,\bm{s})=
\begin{cases}
1, & \text{if } \bm{l}=\bm{l}_0 \\
0, & \text{otherwise}   
\end{cases}
\end{equation}
such that 
\begin{equation}
\begin{split}
&\mathcal{L}_0=\sum_{\bm{s},\bm{l}_0,\bm{l}_1}p(\bm{l}_1,\bm{l}_0,\bm{s})\log\frac{1}{q_{\bm{\theta}}(\bm{l}_0|\bm{l}_1,\bm{s})}  \\
&=\sum_{\bm{s},\bm{l}_0,\bm{l}_1,\bm{l}}p(\bm{l}_1,\bm{l}_0,\bm{s})p(\bm{l}|\bm{l}_1,\bm{l}_0,\bm{s})\log\frac{p(\bm{l}|\bm{l}_1,\bm{l}_0,\bm{s})}{q_{\bm{\theta}}(\bm{l}|\bm{l}_1,\bm{s})} \\
&=\sum_{\bm{s},\bm{l}_0,\bm{l}_1}p(\bm{l}_1,\bm{l}_0,\bm{s})D_{\text{KL}}(p(\bm{l}|\bm{l}_1,\bm{l}_0,\bm{s})||q_{\bm{\theta}}(\bm{l}|\bm{l}_1,\bm{s}))
\end{split}
\end{equation}

The loss function can thus be reduced as 
\begin{equation}
\mathcal{L}(\bm{\theta})=\sum_{t=0}^{T-1}\mathcal{L}_t(\bm{\theta})    
\end{equation}
To ensure practical applicability, the diffusion protocol must satisfy the following two requirements~\cite{austin2021structured}:
\begin{itemize}
\item[(a)] The intermediate state $\bm{l}_t$ can be efficiently sampled from the forward process $p(\bm{l}_t|\bm{l}_0,\bm{s})$;
\item[(b)] The reverse-step distribution $p(\bm{l}_t|\bm{l}_{t+1},\bm{l}_0,\bm{s})$ admits a tractable, closed-form expression.
\end{itemize}

In what follows, we present two concrete diffusion protocols --- continuous diffusion with Gaussian noise and discrete diffusion with random masking --- and demonstrate that both satisfy the above requirements.

In the continuous diffusion setting, $\bm{l}_t$ lies in the continuous space $\mathbb{R}^{n_l}$, and discrete summations over $\bm{l}_t$ are replaced by integrations. The forward process adds independent and identically distributed Gaussian noise to each element of $\bm{l}_{t-1}$ at each step::
\begin{equation}
\label{eq:onestepforward}
\bm{l}_t=\sqrt{1-\beta_t}\bm{l}_{t-1}+\sqrt{\beta_t}\bm{\epsilon}, \quad \epsilon_j\sim\mathcal{N}(0,1)    
\end{equation}
This induces the transition probability
\begin{equation}
p(\bm{l}_t|\bm{l}_{t-1},\bm{l}_0,\bm{s})=p(\bm{l}_t|\bm{l}_{t-1})=\mathcal{N}(\bm{l}_t;\sqrt{1-\beta_t}\bm{l}_{t-1},\beta_t \textbf{I})  
\end{equation}
By recursively substituting $\bm{l}_{t-1}$ into Eq.~\eqref{eq:onestepforward} and noting that the sum of independent Gaussian noises remains Gaussian, we obtain
\begin{equation}
\label{eq:froml0tolt}
\bm{l}_t=\sqrt{\bar{\alpha}_t}\bm{l}_0+\sqrt{1-\bar{\alpha}_t}\bm{\epsilon}, \quad \epsilon_j\sim\mathcal{N}(0,1)
\end{equation}
and the corresponding distribution
\begin{equation}
\label{eq:Probfroml0tolt}
p(\bm{l}_t|\bm{l}_0,\bm{s})=\mathcal{N}(\bm{l}_t;\sqrt{\bar{\alpha}_t}\bm{l}_0,(1-\bar{\alpha}_t)\textbf{I}) 
\end{equation}
where $\alpha_t=1-\beta_t$ and $\bar{\alpha}_t=\prod_{t'=1}^t\alpha_{t'}$. This closed-form transition probability ensures that requirement (a) is met and motivates the use of $q(\bm{l}_T)=\mathcal{N}(\bm{l}_T;\bm{0},\textbf{I})$ as the prior distribution given $\bar{\alpha}_T$ is sufficiently small.

The reverse transition $p(\bm{l}_{t}|\bm{l}_{t+1},\bm{l}_0,\bm{s})$ is Gaussian, obtained via Bayes' rule:
\begin{equation}
\begin{split}
p(\bm{l}_{t-1}|\bm{l}_t,\bm{l}_0,\bm{s})&=\frac{p(\bm{l}_t|\bm{l}_{t-1},\bm{l}_0,\bm{s})p(\bm{l}_{t-1}|\bm{l}_0,\bm{s})}{p(\bm{l}_t|\bm{l}_0,\bm{s})} \\
&=\mathcal{N}(\bm{l}_{t-1};\Tilde{\bm{\mu}}_t(\bm{l}_t,\bm{l}_0), \Tilde{\beta}_t\textbf{I})
\end{split}
\end{equation}
where 
\begin{equation}
\label{eq:true_mean_from_l0}
\Tilde{\bm{\mu}}_t(\bm{l}_t,\bm{l}_0)=\frac{\sqrt{\alpha_t}(1-\bar{\alpha}_{t-1})}{1-\bar{\alpha}_t}\bm{l}_t+\frac{\sqrt{\bar{\alpha}_{t-1}}\beta_t}{1-\bar{\alpha}_t}\bm{l}_0    
\end{equation}
and 
\begin{equation}
\Tilde{\beta}_t=\frac{1-\bar{\alpha}_{t-1}}{1-\bar{\alpha}_t}\beta_t    
\end{equation}
Thus, we can parameterize the learned reverse transition probability $q_{\bm{\theta}}(\bm{l}_{t}|\bm{l}_{t+1},\bm{s})$ also as Gaussian:
\begin{equation}
q_{\bm{\theta}}(\bm{l}_{t-1}|\bm{l}_t,\bm{s})=\mathcal{N}(\bm{l}_{t-1};\bm{\mu}_{\bm{\theta}}(\bm{l}_t,\bm{s},t),\Tilde{\beta}_t\textbf{I})
\end{equation}
Applying the closed-form KL divergence between two Gaussian distributions, we can further simplify $\mathcal{L}_{t}$ as the difference between their means
\begin{equation}
\label{eq:lossMeanDiff}
\begin{split}
&\mathcal{L}_{t-1}(\bm{\theta})=\mathbb{E}_{\bm{s},\bm{l}_0,\bm{l}_t}D_{\text{KL}}(p(\bm{l}_{t-1}|\bm{l}_t,\bm{l}_0,\bm{s})||q_{\bm{\theta}}(\bm{l}_{t-1}|\bm{l}_t,\bm{s}))\\
&=\mathbb{E}_{\bm{s},\bm{l}_0,\bm{l}_t}\frac{1}{2\Tilde{\beta}_t}||\bm{\mu}_{\bm{\theta}}(\bm{l}_t,\bm{s},t)-\Tilde{\bm{\mu}}_t(\bm{l}_t,\bm{l}_0)||^2
\end{split}
\end{equation}
For $\mathcal{L}_0$, Eq.~\eqref{eq:reconstruct_prob_add} can be approximated as a Gaussian distribution $\lim_{\Tilde{\beta}_0\to0}\mathcal{N}(\bm{l};\bm{l}_0,\Tilde{\beta}_0\textbf{I})$ so that the above arguments also hold for $\mathcal{L}_0$.

Using Eq.~\eqref{eq:froml0tolt}, Eq.~\eqref{eq:true_mean_from_l0} can be equivalently written as 
\begin{equation}
\Tilde{\bm{\mu}}_t(\bm{l}_t,\bm{l}_0)=\frac{1}{\sqrt{\alpha_t}}\bm{l}_t-\frac{1-\alpha_t}{\sqrt{1-\bar{\alpha}_t}\sqrt{\alpha_t}}\bm{\epsilon}(\bm{l}_t,\bm{l}_0)
\end{equation}
then we can take 
\begin{equation}
\bm{\mu}_{\bm{\theta}}(\bm{l}_t,\bm{s},t)=\frac{1}{\sqrt{\alpha_t}}\bm{l}_t-\frac{1-\alpha_t}{\sqrt{1-\bar{\alpha}_t}\sqrt{\alpha_t}}\bm{\epsilon}_{\bm{\theta}}(\bm{l}_t,\bm{s},t)
\end{equation}
and set an easier target $\bm{\epsilon}(\bm{l}_t,\bm{l}_0)$ for neural networks to learn. This further reduces Eq.~\eqref{eq:lossMeanDiff} into 
\begin{equation}
\mathcal{L}_{t-1}(\bm{\theta})=\mathbb{E}_{\bm{s},\bm{l}_0,\bm{l}_t}\frac{\beta_t}{2\alpha_t(1-\bar{\alpha}_{t-1})}||\bm{\epsilon}_{\bm{\theta}}(\bm{l}_t,\bm{s},t)-\bm{\epsilon}(\bm{l}_t,\bm{l}_0)||^2    
\end{equation}
The overall training loss is then
\begin{equation}
\mathcal{L}(\bm{\theta})=\mathbb{E}_{\bm{s},\bm{l}}\sum_{t=1}^{T}\omega_t\mathbb{E}_{\bm{l}_{t}\sim p(\bm{l}_{t}|\bm{l},\bm{s})}||\bm{\epsilon}_{\bm{\theta}}(\bm{l}_t,\bm{s},t)-\bm{\epsilon}(\bm{l}_t,\bm{l}_0)||^2
\end{equation}
where $\omega_t=\frac{\beta_t}{2\alpha_t(1-\bar{\alpha}_{t-1})}$ can be either used to sample $t$ or reset as 1. 

Thus far, we have shown requirement (b) is also met, and provided a tractable loss function for optimization.
Based on it, the training algorithm is outlined as follows. 
\begin{itemize}
\item[(A1)] Sample $(\bm{s}, \bm{l})$ from the training dataset;
\item[(A2)] Sample $t\in\{1,2,\dots,T\}$ either uniformly or from $\omega_t$;
\item[(A3)] Sample $\bm{l}_t$ from $p(\bm{l}_{t}|\bm{l},\bm{s})$ (Eq.~\eqref{eq:Probfroml0tolt}), which is equivalent to sampling $\bm{\epsilon}\sim\mathcal{N}(\bm{0},\textbf{I})$ and obtaining $\bm{l}_t$ from Eq.~\eqref{eq:froml0tolt};
\item[(A4)] Take gradient descent step on $\nabla_{\bm{\theta}}||\bm{\epsilon}_{\bm{\theta}}(\bm{l}_t,\bm{s},t)-\bm{\epsilon}(\bm{l}_t,\bm{l}_0)||^2$;
\item[(A5)] Repeat (A1)-(A4) until convergence.
\end{itemize}
When decoding, since the reverse transition probability is Gaussian, with its maximal probability on the mean value, according to Eq.~\eqref{eq:reverse_markov}, the reserve process is realized by moving $\bm{l}_t$ along the path of $\bm{\mu}_{\bm{\theta}}(\bm{l}_t,\bm{s},t)$, i.e.,
\begin{equation}
\hat{\bm{l}}_{t-1}=\frac{1}{\sqrt{\alpha_t}}\hat{\bm{l}}_t-\frac{1-\alpha_t}{\sqrt{1-\bar{\alpha}_t}\sqrt{\alpha_t}}\bm{\epsilon}_{\bm{\theta}}(\hat{\bm{l}}_t,\bm{s},t)    
\end{equation}
This is to be contrasted with the standard diffusion model where an additional random Gaussian noise is added to $\hat{\bm{l}}_{t-1}$ to increase the diversity of the generated samples.

Now we move on to the masked diffusion decoders. In the forward Markov chain, each element from $\bm{l}_t$ is assumed to evolve independently, i.e.,
\begin{equation}
p(\bm{l}_{t}|\bm{l}_{t-1})=\prod_{k=1}^{n_l}p(l_{t,k}|l_{t-1,k})
\end{equation}
And a transition matrix $\bm{Q}_t\in\mathbb{R}^{3\times3}$ is defined such that
\begin{equation}
p(l_{t,k}=j|l_{t-1,k}=i)=Q_{t,ij}    
\end{equation}
Here index 2 denotes the masking value `$*$'. Meanwhile, we have the following identities:
\begin{equation}
p(l_{t,k}|\bm{l}_0,\bm{s})=p(l_{t,k}|l_{0,k})=\overline{Q}_{t,l_{0,k}l_{t,k}}    
\end{equation}
\begin{equation}
\overline{\bm{Q}}_{t}=\bm{Q}_1\bm{Q}_2\cdots\bm{Q}_t
\end{equation}
\begin{equation}
\label{eq:reverse_masked_cond}
\begin{split}
&p(l_{t-1,k}|l_{t,k},\bm{l}_0,\bm{s})=\frac{p(l_{t,k}|l_{t-1,k},\bm{l}_0,\bm{s})p(l_{t-1,k}|\bm{l}_0,\bm{s})}{p(l_{t,k}|\bm{l}_0,\bm{s})}\\
&=\frac{\overline{Q}_{t-1,l_{0,k}l_{t-1,k}}Q_{t,l_{t-1,k}l_{t,k}}}{\overline{Q}_{t,l_{0,k}l_{t,k}}}
\end{split}    
\end{equation}
Now let us give the expression of $\bm{Q}_t$ by assuming that `$*$' is an absorbing state; in other words, once an element transits to `$*$', it will no longer transit to other states in this forward masking process.
\begin{equation}
Q_{t,ij}=
\begin{cases}
1, & i=*,\;j=* \\
1-\beta_t, & i\neq*,\;j=i \\
\beta_t, & i\neq*,\;j=* 
\end{cases}
\end{equation}
Here $\beta_t$ is the probability of transiting to `$*$' at time step $t$. It can also be written as 
\begin{equation}
\label{eq:Qt}
\bm{Q}_t=(1-\beta_t)I+\beta_t \mathbb{1} e_m^T    
\end{equation}
where $e_m^T=(0,0,1)$ is the one-hot vector for masking value, and $\mathbb{1}=(1,1,1)^T$. Consequently, 
\begin{equation}
\label{eq:Qt_bar}
\overline{\bm{Q}}_{t}=\alpha_tI+(1-\alpha_t)\mathbb{1} e_m^T
\end{equation}
with $\alpha_t=\prod_{t'=1}^t(1-\beta_{t'})$. Notice that each row of $\bm{Q}_t$ and $\overline{\bm{Q}}_t$ only has two non-zero entries, indicating that this element either stays the same as the initial value or gets masked. Now, let us determine $\beta_t$ if we want the expected ratio of masked elements at time step $t$ to be $t/T$. Since each element evolves independently, we get the following equation:
\begin{equation}
\frac{t}{T}=p(l_{t,k}=*|l_{0,k}\neq*)=1-\alpha_t 
\end{equation}
which gives $\alpha_t=1-t/T$, $\beta_t=1/(T-t+1)$.
Under this condition, the prior distribution is simply $q(\bm{l}_T)=1$ for $\bm{l}_T=(*,*,\dots,*)$ and 0 otherwise, and the prior matching term $\mathcal{L}_T$ from Eq.~\eqref{eq:ELBO} vanishes; the requirements (a) and (b) are satisfied as well.  

Since the ground-truth reverse-step transition probability $p(\bm{l}_{t-1}|\bm{l}_t,\bm{l}_0,\bm{s})$ is factorized into $\prod_{k=1}^{n_l}p(l_{t-1,k}|l_{t,k},\bm{l}_0,\bm{s})$, we also factorize the variational reserve-step probability $q_{\bm{\theta}}(\bm{l}_{t-1}|\bm{l}_{t},\bm{s})=\prod_{k=1}^{n_l}q_{\bm{\theta}}(l_{t-1,k}|\bm{l}_{t},\bm{s})$ and parameterize each individual term as 
\begin{equation}
q_{\bm{\theta}}(l_{t-1,k}|\bm{l}_{t},\bm{s})=\sum_{l_{0,k}}p(l_{t-1,k}|\bm{l}_{t},l_{0,k},\bm{s})q_{\bm{\theta}}(l_{0,k}|\bm{l}_{t},\bm{s})
\end{equation}
Notice that from Eq.~\eqref{eq:reverse_masked_cond},
~\eqref{eq:Qt}, ~\eqref{eq:Qt_bar}, we have 
\begin{equation}
\begin{split}
&p(l_{t-1,k}|\bm{l}_{t},l_{0,k},\bm{s})\\
=&
\begin{cases}
1, & l_{t,k}=l_{0,k},\;l_{t-1,k}=l_{t,k}\\[3pt]
\frac{1-\alpha_{t-1}}{1-\alpha_t}=\frac{t-1}{t}, & l_{t,k}=*,\;l_{t-1,k}=*\\[3pt]
\frac{\alpha_{t-1}\beta_t}{1-\alpha_t}=\frac{1}{t}, & l_{t,k}=*,\;l_{t-1,k}=l_{0,k}
\\[3pt]
0, & \text{otherwise}
\end{cases}
\end{split}
\end{equation}
Therefore,
\begin{equation}
q_{\bm{\theta}}(l_{t-1,k}|\bm{l}_{t},\bm{s})
=\begin{cases}
1, & l_{t,k}\neq *,\;l_{t-1,k}=l_{t,k}\\[3pt]
\frac{t-1}{t}, & l_{t,k}=*,\;l_{t-1,k}=* \\[3pt]
\frac{1}{t}q_{\bm{\theta}}(l_{0,k}|\bm{l}_{t},\bm{s}), & l_{t,k}=*,\;l_{t-1,k}=l_{0,k}\\[3pt]
0, & \text{otherwise}
\end{cases}
\end{equation}
In other words, for DF decoders, we only need a time-independent predictor $q_{\bm{\theta}}(l_{0,k}|\bm{l}_{t},\bm{s})$. Next, we substitute the above expressions into $\mathcal{L}_t(\bm{\theta})$, and ignore the constant terms.
\begin{equation}
\mathcal{L}_{t-1}(\bm{\theta})=\mathbb{E}_{\bm{s},\bm{l}_0,\bm{l}_t}\left[-\frac{1}{t}\sum_{\{k|l_{t,k}=*\}}\log q_{\bm{\theta}}(l_{0,k}|\bm{l}_{t},\bm{s})\right]    
\end{equation}
This gives a tractable loss function for optimization. The overall training loss is then
\begin{equation}
\label{eq:lossMaskFinal}
\mathcal{L}(\bm{\theta})=\mathbb{E}_{\bm{s},\bm{l}}\sum_{t=1}^{T}\frac1t\mathbb{E}_{\bm{l}_{t}\sim p(\bm{l}_{t}|\bm{l},\bm{s})}\sum_{\{k|l_{t,k}=*\}}-\log q_{\bm{\theta}}(l_{0,k}|\bm{l}_{t},\bm{s})    
\end{equation}
In practice, since in expectation $n_l\cdot t/T$ elements are masked at time step $t$, and to reduce the estimation variance, the sampling of $t$ can be replaced by randomly selecting $n_l\cdot t/T$ to be masked. 
From this tractable loss function, the training algorithm is outlined as follows: 
\begin{itemize}
\item[(B1)] Sample $(\bm{s}, \bm{l})$ from the training dataset;
\item[(B2)] Sample $t\in\{1,2,\dots,T\}$ uniformly;
\item[(B3)] Sample $\bm{l}_t$ from $p(\bm{l}_{t}|\bm{l},\bm{s})$, which is equivalent to randomly selecting $n_l\cdot t/T$ elements from $\bm{l}$ without replacement to be masked;
\item[(B4)] Take gradient descent step on $\nabla_{\bm{\theta}}\left(-\frac1t\right)\sum_{\{k|l_{t,k}=*\}}\log q_{\bm{\theta}}(l_{0,k}|\bm{l}_{t},\bm{s})$;
\item[(B5)] Repeat (B1)-(B4) until convergence.
\end{itemize}

When decoding, at each time step $t$, in expectation $n_l\cdot t/T$ elements should remain masked; To obey Eq.~\eqref{eq:reverse_markov}, we choose the most confident $n_l/T$ elements --- those with largest $q_{\bm{\theta}}(l_{0,k}|\bm{l}_{t},\bm{s})$ --- to unmask at one step. 

\bibliography{main}
\end{document}